\documentclass[12pt,english,12pt,onecolumn, draftcls]{IEEEtran}
\usepackage[latin9]{inputenc}
\usepackage{amsmath}
\usepackage{amssymb}
\usepackage{graphicx}

\makeatletter
\usepackage{amsfonts}
\usepackage[T1]{fontenc}
\usepackage[latin9]{inputenc}
\usepackage{units}
\usepackage{amsthm}
\usepackage{amsmath}
\usepackage{amssymb}
\usepackage{graphicx}
\usepackage{esint}
\PassOptionsToPackage{version=3}{mhchem}
\usepackage{mhchem}
\providecommand{\U}[1]{\protect\rule{.1in}{.1in}}
\newtheorem{theorem}{Theorem}

\newtheorem{proposition}{Proposition}
\newtheorem{remark}[theorem]{Remark}

\theoremstyle{definition}
\newtheorem{definition}{Definition}

\makeatother

\begin{document}

\title{Distributed and Cascade Lossy Source Coding with a Side Information
``Vending Machine''}

\author{Behzad Ahmadi and Osvaldo Simeone%
\thanks{This work has been supported by the U.S. National Science Foundation
under grant CCF-0914899.%
} \and Department of Electrical and Computer Engineering \and New
Jersey Institute of Technology \and University Heights, Newark, New
Jersey 07102 \and Email: behzad.ahmadi@njit.edu, osvaldo.simeone@njit.edu}

\author{\authorblockN{Behzad Ahmadi and Osvaldo Simeone}\\
\thanks{This work has been supported by the U.S. National Science Foundation
under grant CCF-0914899.%
} \authorblockA{Department of Electrical and Computer Engineering\\
 New Jersey Institute of Technology\\
 University Heights, Newark, New Jersey 07102\\
 Email: behzad.ahmadi@njit.edu, osvaldo.simeone@njit.edu}}
\maketitle
\begin{abstract}
Source coding with a side information ``vending machine'' is a recently
proposed framework in which the statistical relationship between the
side information and the source, instead of being given and fixed
as in the classical Wyner-Ziv problem, can be controlled by the decoder.
This control action is selected by the decoder based on the message
encoded by the source node. Unlike conventional settings, the message
can thus carry not only information about the source to be reproduced
at the decoder, but also control information aimed at improving the
quality of the side information.

In this paper, the analysis of the trade-offs between rate, distortion
and cost associated with the control actions is extended from the
previously studied point-to-point set-up to two basic multiterminal
models. First, a distributed source coding model is studied, in which
two encoders communicate over rate-limited links to a decoder, whose
side information can be controlled. The control actions are selected
by the decoder based on the messages encoded by both source nodes.
For this set-up, inner bounds are derived on the rate-distortion-cost
region for both cases in which the side information is available causally
and non-causally at the decoder. These bounds are shown to be tight
under specific assumptions, including the scenario in which the sequence
observed by one of the nodes is a function of the source observed
by the other and the side information is available causally at the
decoder. Then, a cascade scenario in which three nodes are connected
in a cascade and the last node has controllable side information,
is also investigated. For this model, the rate-distortion-cost region
is derived for general distortion requirements and under the assumption
of causal availability of side information at the last node.

\textmd{$\textbf{Keywords}$}: Distributed source coding, cascade
source coding, observation costs, side information, side information
vending machine, rate-distortion theory.
\end{abstract}

\section{Introduction}

\begin{figure}[h!]
\centering\includegraphics[bb=112bp 433bp 402bp 749bp,clip,scale=0.65]{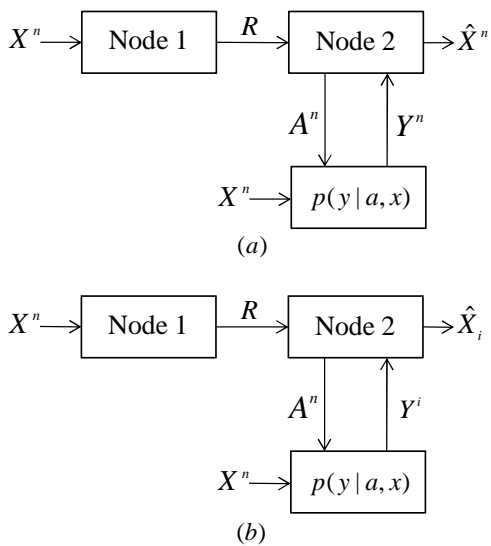}
\caption{Source coding with a vending machine at the decoder \cite{Permuter}
with: (a) ``non-causal'' side information; (b) ``causal'' side
information.}

\label{fig:fig0} 
\end{figure}

Reference \cite{Permuter} introduced the notion of a side information
``vending machine''. To illustrate the idea, consider the setting
in Fig. \ref{fig:fig0}, as studied in \cite{Permuter}. Here, unlike
the conventional Wyner-Ziv set-up (see, e.g., \cite[Chapter 12]{Elgammal}),
the joint distribution of the side information $Y$ available at the
decoder (Node 2) and of the source $X$ observed at the encoder (Node
1) is not given. Instead, it can be controlled through the selection
of an ``action'' $A$, so that, for a given action $A$ and source
symbol $X$, the side information $Y$ is distributed according to
a given conditional distribution $p(y|a,x).$ Action $A$ is selected
by the decoder based on the message $M$, of $R$ bits per source
symbol, received from the encoder, and is subject to a cost constraint.
The latter limits the ``quality'' of the side information that can
be collected by the decoder.

The source coding problem with a vending machine provides a useful
model for scenarios in which acquiring data as side information is
costly and thus should be done effectively. Examples include computer
networks, in which data must be obtained from remote data bases, and
sensor networks, where data is acquired via measurements.

The key aspect of this model is that the message $M$ produced by
the encoder plays a double role. In fact, on the one hand, it needs
to carry the description of the source $X$ itself, as in, e.g., the
standard Wyner-Ziv model. On the other hand, it can also carry \textit{control}
information aimed at enabling the decoder to make an appropriate selection
of action $A.$ The goal of such a selection is to obtain a side information
$Y$ that is better suited to provide partial information about the
source $X$ to the decoder. This in turn can potentially reduce the
rate $R$ necessary for the decoder to reconstruct source $X$ at
a given distortion level (or, vice versa, to reduce the distortion
level for a given rate $R$).

The performance of the system in Fig. \ref{fig:fig0} is expressed
in terms of the interplay among three metrics, namely the rate $R$,
the cost budget $\Gamma$ on the action $A,$\ and the distortion
$D$ of the reconstruction $\hat{X}$ at the decoder. This trade-off
is summarized by the \textit{rate-distortion-cost} function $R(D,\Gamma).$
This function characterizes the infimum of all rates $R$ for which
a distortion level $D$ can be achieved under an action cost budget
$\Gamma,$ by allowing encoding of an arbitrary number $n$ of source
symbols $X^{n}=(X_{1},...,X_{n}).$ This function is derived in \cite{Permuter}
for both cases in which the side information $Y$ is available ``non-causally''
to the decoder, as in the standard Wyner-Ziv model, or ``causally'',
as introduced in \cite{Elgammal Weissman}. In the former case (Fig.
\ref{fig:fig0}-(a)), the estimated sequence $\hat{X}^{n}=(\hat{X}_{1},...,\hat{X}_{n})$
is a function of message $M$ and of the entire side information sequence
$Y^{n}=(Y_{1},...,Y_{n})$, while, in the latter (Fig. \ref{fig:fig0}-(b)),
each estimated sample $\hat{X}_{i}$ is a function of message $M$
and the side information as received up to time $i$, i.e., $Y^{i}=(Y_{1},...,Y_{i})$
for $i=1,...,n$. We note that the model with causal side information
is appropriate, for instance, when there are delay constraints on
the reproduction at the decoder or when the decoder operates by filtering
the side information sequence. We refer to \cite[Sec I]{Elgammal Weissman}
for an extensive discussion on these points.

Following reference \cite{Permuter}, recent works \cite{Ahmadi}
and \cite{Weissman multi} generalized the characterization of the
rate-distortion-cost function for the models in Fig. \ref{fig:fig0}
to a set-up analogous to the so called Kaspi-Heegard-Berger problem
\cite{Heegard and Berger}\cite{Kaspi}, in which the side information
vending machine may or may not be available at the decoder. This entails
the presence of \textit{two decoders}, rather than only one as in
Fig. \ref{fig:fig0}, one with access to the vending machine and one
without any side information. Reference \cite{Ahmadi,Weissman multi}
also solved the more general case in which both decoders have access
to the same vending machine, and either the side informations produced
by the vending machine at the two decoders satisfy a degradedness
condition, or lossless source reconstructions are required at the
decoders. The papers \cite{Tobias}\cite{Tobias2} studied the setting
of Fig. \ref{fig:fig0} but under the additional constraints of common
reconstruction, in the sense of \cite{Steinberg}, in \cite{Tobias},
and of secrecy with respect to an ``eavesdropping'' node in \cite{Tobias2},
providing characterizations of the corresponding achievable performance.
The impact of actions that adapt to the previously measured samples
of the side information is studied in \cite{Chiru}. Finally, real-time
constraints are investigated in \cite{Asnani real time}.

\subsection{Contributions and Overview}

\begin{figure}[h!]
\centering\includegraphics[bb=68bp 538bp 421bp 702bp,clip,scale=0.65]{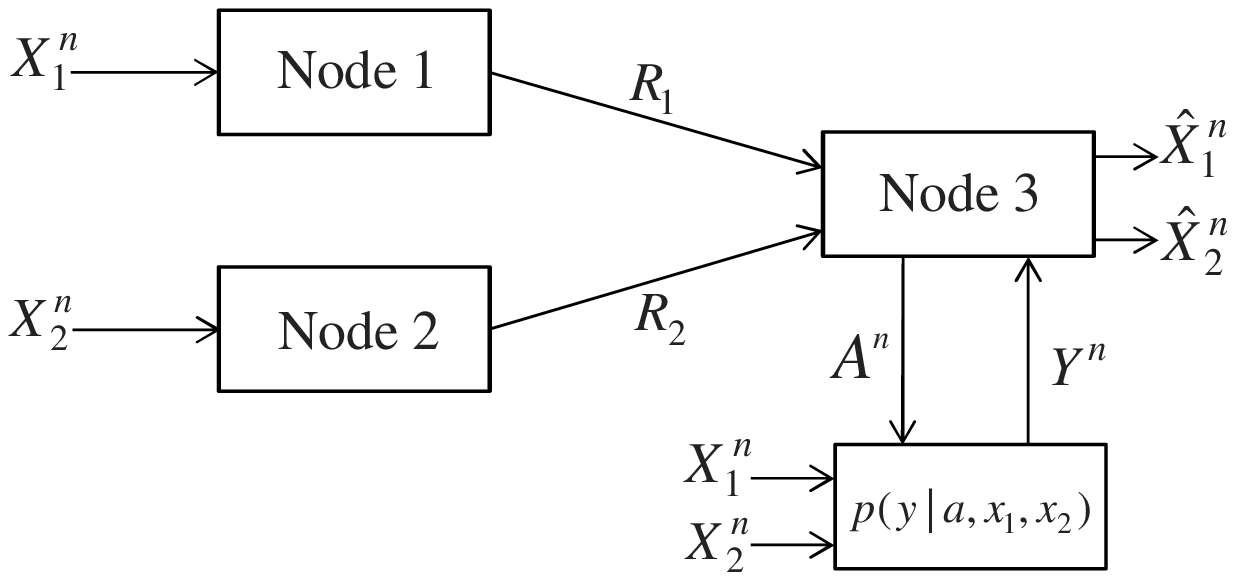}
\caption{Distributed source coding with a side information vending machine
at the decoder.}

\label{fig:fig1} 
\end{figure}

\begin{figure}[h!]
\centering\includegraphics[bb=64bp 520bp 507bp 690bp,clip,scale=0.65]{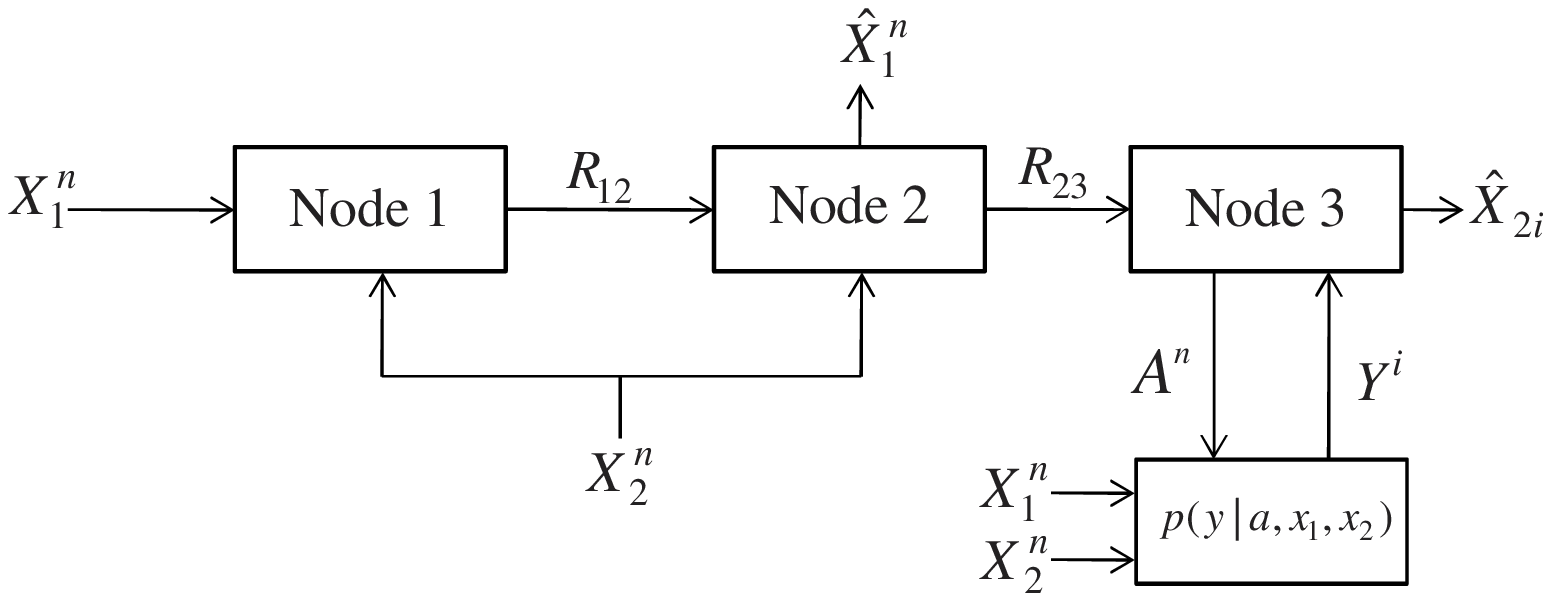}
\caption{Cascade source coding with a side information vending machine. Side
information is assumed to be available ``causally'' to the decoder.}

\label{fig:fig3} 
\end{figure}

In this paper, we study two multi-terminal extensions of the set-up
in Fig. \ref{fig:fig0}, namely the \textit{distributed source coding}
setting of Fig. \ref{fig:fig1}, and the \emph{cascade} model of Fig.
\ref{fig:fig3}. The analysis of these scenarios is motivated by the
observation that they constitute key components of computer and sensor
networks. In fact, as discussed above, an important aspect of these
networks is the need to effectively acquire side information data,
which can be modeled by including a side information vending machine.
We overview the two extensions and the corresponding main results
below.

1) \textit{Distributed source coding with a side information vending
machine} (Sec. \ref{sec:distsc}): In the distributed source coding
setting of Fig. \ref{fig:fig1}, \textit{two encoders }(Node 1 and
Node 2)\textit{,} which measure correlated sources\textit{ }$X_{1}$\textit{
}and $X_{2}$\textit{,} respectively, communicate over rate-limited
links, of rates $R_{1}$ and $R_{2}$, respectively, to a single decoder
(Node 3). The decoder has side information $Y$ on sources $X_{1}$
and $X_{2},$ which can be controlled through an action $A.$ The
action sequence is selected by the decoder based on the messages $M_{1}$
and $M_{2}$ received from Node 1 and Node 2, respectively, and needs
to satisfy a cost constraint of $\Gamma$. Inner bounds are derived
to the rate-distortion-cost region ${\cal R}(D_{1},D_{2},\Gamma)$
under non-causal and causal side information by combining the strategies
proposed in \cite{Permuter} with the Berger-Tung strategy \cite{Tung}
and its extension to the Wyner-Ziv set-up \cite{Gastpar}. These bounds
are shown to be tight under specific assumptions, including the scenario
where the sequence observed by one of the nodes is a function of the
source observed by the other and the side information is available
causally at the decoder.

2) \textit{Cascade source coding with a side information vending machine}
(Sec. \ref{sec:cascadesc}):\textbf{ }In the cascade model of Fig.
\ref{fig:fig3}, Node 1 is connected via a rate-limited link, of rate
$R_{12}$, to Node 2, which is in turn communicates with Node 3 with
rate $R_{23}$. Source $X_{1}$ is measured by Node 1 and the correlated
source $X_{2}$ by both Node 1 and Node 2. Similarly to the distributed
coding setting described above, Node 3 has side information $Y$ on
sources $X_{1}$ and $X_{2},$ which can be controlled via an action
$A$. Action $A$ is selected by Node 3 based on the message received
from Node 2 and needs to satisfy a cost constraint of $\Gamma.$ We
derive the set $\mathcal{R}(D_{1},D_{2},\Gamma)$ of all achievable
rates ($R_{12},R_{23}$) for given distortion constraints ($D_{1},D_{2})$
on the reconstructions $\hat{X}_{1}$ and $\hat{X}_{2}$ at Node 2
and Node 3, respectively, and for cost constraint $\Gamma.$ This
characterization is obtained under the assumption that the side information
$Y$ be available causally at Node 3. It is mentioned that, following
the submission of this work, the analysis of the case with non-causal
side information at Node 3 was carried out in \cite{Ahmadi_Chiru}.

\textit{Notation}: For $a,b$ integer with $a\leq b$, we define $[a,b]$
as the interval $[a,a+1,...,b]$ and $x_{a}^{b}=(x_{a},...,x_{b})$;
if instead $a>b$ we set $[a,b]=\emptyset$ and $x_{a}^{b}=\emptyset$.
We will also write $x_{1}^{b}$ for $x^{b}$ for simplicity of notation.
Random variables are denoted with capital letters and corresponding
values with lowercase letters. Given random variables, or more generally
vectors, $X$ and $Y,$ we will use the notation $p_{X}(x)$ or $p(x)$
for $\Pr[X=x]$, and $p_{X|Y}(x|y)$ or $p(x|y)$ for $\Pr[X=x|Y=y]$,
where the latter notations are used when the meaning is clear from
the context. Given set $\mathcal{X}$, we define as $\mathcal{X}^{n}$
the $n$-fold Cartesian product of $\mathcal{X}$. Function $\delta(x)$
represents the Kronecker delta function, i.e., $\delta(x)=1$ if $x=0$
and $\delta(x)=0$ otherwise.

\section{Distributed Source Coding with a Side Information Vending Machine
\label{sec:distsc}}

In this section, we first detail the system model for the problem
of distributed source coding with a side information vending machine
in Sec. \ref{model_general}. Then, we propose an achievable strategy
in Sec. \ref{sub:Achievable-Strategy} for both the cases with non-causal
and causal side information at the decoder. In Sec. \ref{sec:XY,Y-Causal}
and Sec. \ref{sec:distsc-X,Y-NonCausal-lossy-lossless} scenarios
are discussed in which the achievable strategies match given outer
bounds. A numerical example is then developed in Sec. \ref{example}.

\subsection{System Model\label{model_general}}

The problem of distributed lossy source coding with a vending machine
and non-causal side information is illustrated in Fig. \ref{fig:fig1}.
It is defined by the probability mass functions (pmfs) $p_{X_{1}X_{2}}(x_{1},x_{2})$
and $p_{Y|AX_{1}X_{2}}(y|a,x_{1},x_{2})$ and discrete alphabets $\mathcal{X}_{1},\mathcal{X}_{2},\mathcal{Y},\mathcal{A},\mathcal{\hat{X}}_{1},\mathcal{\hat{X}}_{2}$
as follows. The source sequences $X_{1}^{n}$ and $X_{2}^{n}$ with
$X_{1}^{n}\in\mathcal{X}_{1}^{n}$ and $X_{2}^{n}\in\mathcal{X}_{2}^{n}$,
respectively, are such that the tuples $(X_{1i},X_{2i})$ for $i\in[1,n]$
are independent identically distributed (i.i.d.) with joint pmf $p_{X_{1}X_{2}}(x_{1},x_{2})$.
Node 1 measures sequences $X_{1}^{n}$ and encodes it into message
$M_{1}$ of $nR_{1}$ bits, while Node 2 measures sequences $X_{2}^{n}$
and encodes it into message $M_{2}$ of $nR_{2}$ bits. Node 3 wishes
to reconstruct the two sources within given distortion requirements,
to be discussed below, as $\hat{X}_{1}^{n}\in\mathcal{\hat{X}}_{1}^{n}$
and $\hat{X}_{2}^{n}\in\mathcal{\hat{X}}_{2}^{n}$.

To this end, Node 3 selects an action sequence $A^{n},$ where $A^{n}\in\mathcal{A}^{n},$
based on the messages $M_{1}$ and $M_{2}$ received from Node 1 and
Node 2, respectively. The side information sequence $Y^{n}$ is then
realized as the output of a memoryless channel with inputs ($A^{n},X_{1}^{n},X_{2}^{n}$).
Specifically, given $A^{n}$, $X_{1}^{n}$ and $X_{2}^{n}$, the sequence
$Y^{n}$ is distributed as
\begin{equation}
p(y^{n}|a^{n},x_{1}^{n},x_{2}^{n})=\underset{i=1}{\overset{n}{\prod}}p_{Y|AX_{1}X_{2}}(y_{i}|a_{i},x_{1i},x_{2i}).\label{pzxya}
\end{equation}
The overall cost of an action sequence $a^{n}$ is defined by a per-symbol
cost function $\Lambda$: $\mathcal{A\rightarrow}[0,\Lambda_{\max}]$
with $0\leq\Lambda_{\max}<\infty,$ as 
\begin{equation}
\Lambda^{n}(a^{n})=\frac{1}{n}\sum_{i=1}^{n}\Lambda(a_{i}).\label{overall_cost}
\end{equation}
The estimated sequences $\hat{X}_{1}^{n}$ and $\hat{X}_{2}^{n}$
are obtained as a function of both messages $M_{1}$ and $M_{2}$
and of the side information $Y$. The estimates $\hat{X}_{1}^{n}$
and $\hat{X}_{2}^{n}$ are constrained to satisfy distortion constraints
defined by two per-symbol distortion measures, namely $d_{j}(x_{1},x_{2},y,\hat{x}_{j})$:
$\mathcal{X}_{1}\times\mathcal{X}_{2}\times\mathcal{Y}\times\mathcal{\hat{X}}_{j}\rightarrow[0,D_{\max}]$
for $j=1,2$ with $0\leq D_{\max}<\infty$. Based on such scalar measures,
the overall distortion for the estimated sequences $\hat{x}_{1}^{n}$
and $\hat{x}_{2}^{n}$ is defined as 
\begin{align}
d_{j}^{n}(x_{1}^{n},x_{2}^{n},y^{n},\hat{x}_{j}^{n}) & =\frac{1}{n}\sum_{i=1}^{n}d_{j}(x_{1i},x_{2i},y_{i},\hat{x}_{ji})\mbox{ for }j=1,2.\text{ }\label{eq:per-letter}
\end{align}
Note that, based on (\ref{eq:per-letter}), the estimate $\hat{X}_{j}^{n}$
$\mbox{ for }j=1,2$ can be required to be a lossy version of an arbitrary
(per-letter) function of both sources $X_{1}^{n}$ and $X_{2}^{n}$
and of the side information sequence $Y^{n}$. A formal description
of the operations at encoders and decoder, and of cost and distortion
constraints, is presented below for both the cases in which the side
information is available causally or non-causally at the decoder.

\begin{definition} \label{def1}An $(n,R_{1},R_{2},D_{1},D_{2},\Gamma)$
code for the case of \emph{non-casual} side information at Node 3
consists of two source encoders
\begin{align}
\mathrm{g}_{1} & \text{:}\text{ }\mathcal{X}_{1}^{n}\rightarrow[1,2^{nR_{1}}],\nonumber \\
\text{and }\mathrm{g}_{2} & \text{:}\text{ }\mathcal{X}_{2}^{n}\rightarrow[1,2^{nR_{2}}],
\end{align}
which map the sequences $X_{1}^{n}$ and $X_{2}^{n}$ into messages
$M_{1}$ and $M_{2}$ at Node 1 and Node 2, respectively; an ``action\textquotedblright{}\ function
\begin{equation}
\mathrm{\ell}\text{: }[1,2^{nR_{1}}]\times[1,2^{nR_{2}}]\rightarrow\mathcal{A}^{n},\label{action enc}
\end{equation}
which maps the message $(M_{1},M_{2})$ into an action sequence $A^{n}$
at Node 3; and two decoding functions
\begin{align}
\mathrm{h}_{1} & \text{:}\text{ }[1,2^{nR_{1}}]\times[1,2^{nR_{2}}]\times{\cal Y}^{n}\rightarrow\mathcal{\hat{X}}_{1}^{n},\label{eq:dec1_NC}\\
\text{and }\mathrm{h}_{2} & \text{:}\text{ }[1,2^{nR_{1}}]\times[1,2^{nR_{2}}]\times{\cal Y}^{n}\rightarrow\mathcal{X}_{2}^{n},\label{eq:dec2_NC}
\end{align}
which map the messages $M_{1}$ and $M_{2}$, and the side information
sequence $Y^{n}$ into the estimated sequences $\hat{X}_{1}^{n}$
and $\hat{X}_{2}^{n}$ at Node 3; such that the action cost constraint
$\Gamma$ is satisfied as
\begin{equation}
\frac{1}{n}\underset{i=1}{\overset{n}{\sum}}\mathrm{E}\left[\Lambda(A_{i})\right]\leq\Gamma,\label{action_cost_const}
\end{equation}
and the distortion constraints $D_{1}$ and $D_{2}$ hold, namely
\begin{align}
\frac{1}{n}\underset{i=1}{\overset{n}{\sum}}\mathrm{E}\left[d_{j}(X_{1i},X_{2i},Y_{i},\hat{X}_{ji})\right] & \leq D_{j},\mbox{ for }j=1,2.\label{eq:dist_const}
\end{align}
\end{definition}

\begin{definition} \label{def1_C} A $(n,R_{1},R_{2},D_{1},D_{2},\Gamma)$
code for the case of \emph{causal} side information at Node 3 is as
in Definition \ref{def1} with the only difference that, in lieu of
(6)-(7), we have the sequence of decoding functions 
\begin{align}
\mathrm{h}_{1i} & \text{:}\text{ }[1,2^{nR_{1}}]\times[1,2^{nR_{2}}]\times{\cal Y}^{i}\rightarrow\mathcal{\hat{X}}_{1i},\label{eq:dec1_C}\\
\text{and }\mathrm{h}_{2i} & \text{:}\text{ }[1,2^{nR_{1}}]\times[1,2^{nR_{2}}]\times{\cal Y}^{i}\rightarrow\mathcal{X}_{2i},\label{eq:dec2_C}
\end{align}
for $i\in[1,n]$, which map the message $(M_{1},M_{2})$ and the measured
sequence $Y^{i}$ into the $i$th estimated symbol $\hat{X}_{ji}=\mathrm{h}_{ji}(M_{1},M_{2},Y^{i})$
for $j=1,2$ at Node 3.\end{definition}

\begin{definition}\label{def2}Given a distortion-cost tuple $(D_{1},D_{2},\Gamma)$,
a rate pair $(R_{1},R_{2})$ is said to be achievable for the case
with non-causal or causal side information if, for any $\epsilon>0$
and sufficiently large $n$, there exists a corresponding $(n,R_{1},R_{2},D_{1}+\epsilon,D_{2}+\epsilon,\Gamma+\epsilon)$
code.\end{definition}

\begin{definition} \label{def3}The \textit{rate-distortion-cost
region }$\mathcal{R}_{NC}(D_{1},D_{2},\Gamma)$ is defined as the
closure of all rate pairs $(R_{1},R_{2})$ that are achievable with
non-causal side information given the distortion-cost tuple $(D_{1},D_{2},\Gamma)$.
The rate-distortion-cost region $\mathcal{R}_{C}(D_{1},D_{2},\Gamma)$
is similarly defined for the case of casual side information.\end{definition}

\subsection{Achievable Strategies\label{sub:Achievable-Strategy}}

In this section, we obtain inner bounds to the rate-distortion-cost
regions for the cases with non-causal and causal side information.

\begin{proposition}\label{prop:ach_NC} The rate-distortion-cost
region with non-causal side information at Node 3 satisfies the inclusion
${\cal R}_{NC}(D_{1},D_{2},\Gamma)\supseteq{\cal R}_{NC}^{a}(D_{1},D_{2},\Gamma)$,
where the region ${\cal R}_{NC}^{a}(D_{1},D_{2},\Gamma)$ is given
by the union of the set of all of rate tuples $(R_{1},R_{2})$ that
satisfy the inequalities\emph{ }\begin{subequations}\emph{\label{DSC_ach_reg_NC}}
\begin{align}
R_{1} & \geq I(X_{1};V_{1}|V_{2},Q)+I(X_{1};U_{1}|V_{1},V_{2},U_{2},Y,Q)\label{ach_R1_NC}\\
R_{2} & \geq I(X_{2};V_{2}|V_{1},Q)+I(X_{2};U_{2}|V_{1},V_{2},U_{1},Y,Q)\label{ach_R2_NC}\\
\text{and }R_{1}+R_{2} & \geq I(X_{1},X_{2};V_{1},V_{2}|Q)+I(X_{1},X_{2};U_{1},U_{2}|V_{1},V_{2},Y,Q),\label{ach_R1+R2_NC}
\end{align}
for some joint pmfs that factorizes as\end{subequations}
\begin{eqnarray}
p(q,x_{1},x_{2},y,v_{1},v_{2},u_{1},u_{2},a,\hat{x}_{1},\hat{x}_{2})\negmedspace\negmedspace\negmedspace & = & \negmedspace\negmedspace\negmedspace p(q)p(x_{1},x_{2})p(v_{1},u_{1}|x_{1},q)p(v_{2},u_{2}|x_{2},q)\delta(a-\mathrm{a}(v_{1},v_{2},q))\nonumber \\
\negmedspace\negmedspace &  & \negmedspace\negmedspace\negmedspace p(y|a,x_{1},x_{2})\delta(\hat{x}_{1}-\mathrm{\hat{x}}_{1}(u_{1},u_{2},y,q))\nonumber \\
\negmedspace\negmedspace &  & \negmedspace\negmedspace\negmedspace\delta(\hat{x}_{2}-\mathrm{\hat{x}}_{2}(u_{1},u_{2},y,q)),\label{eq:joint_ach_NC}
\end{eqnarray}
with pmfs $p(q)$ and $p(v_{1},u_{1}|x_{1},q)$ and $p(v_{2},u_{2}|x_{2},q)$
and deterministic functions $\mathrm{a}\textrm{\ensuremath{\mathrm{:}} }\mathcal{V}_{1}\times\mathcal{V}_{2}\times{\cal Q}\rightarrow\mathcal{A}$,
$\mathrm{\hat{x}}_{j}\textrm{\ensuremath{\mathrm{:}} }\mathcal{U}_{1}\times\mathcal{U}_{2}\times\mathcal{Y}\times Q\rightarrow\hat{\mathcal{X}}_{j}$
for $j=1,2$, such that the action and the distortion constraints\begin{subequations}\label{DSC_General_const}
\begin{align}
\mathrm{E}\left[\Lambda(A)\right] & \leq\Gamma\label{eq:action_const_ach}\\
\text{and }\mathrm{E}\left[d_{j}(X_{1},X_{2},Y,\hat{X}_{j})\right] & \leq D_{j},\mbox{ for }j=1,2,\label{eq:dist_const_ach}
\end{align}
\end{subequations}hold. Finally, any extreme point of the region
${\cal R}_{NC}^{a}(D_{1},D_{2},\Gamma)$ can be obtained by limiting
the cardinalities of the random variables $(V_{1},V_{2},U_{1},U_{2})$
as $\left\vert \mathcal{V}_{j}\right\vert \leq\left\vert \mathcal{X}_{j}\right\vert +6$
and $\left\vert \mathcal{U}_{j}\right\vert \leq\left\vert \mathcal{X}_{j}\right\vert \left\vert \mathcal{V}_{j}\right\vert +5$,
for $j=1,2$.\end{proposition}

\begin{remark} If we set $p(y|a,x_{1},x_{2})=p(y|x_{1},x_{2}),$
so that the side information is action-independent, Proposition \ref{prop:ach_NC}
reduces to the extension of the Berger-Tung scheme \cite{Tung} to
the Wyner-Ziv set-up studied in \cite[Theorem 2]{Gastpar}. Moreover,
in the special case in which there is only one encoder, the achievable
rate coincides with that derived in \cite[Theorem 1]{Permuter}. \end{remark}

The proof of Proposition \ref{prop:ach_NC} follows easily from standard
arguments, and thus it is only briefly discussed here. The proposed
scheme combines the Berger-Tung distributed source coding strategy
\cite{Tung} and the distributed Wyner-Ziv approach proposed in \cite[Theorem II]{Gastpar}
with the layered two-stage coding scheme that is proved to be optimal
in \cite{Permuter} for the special case of a single encoder. Throughout
the discussion we neglect the time-sharing variable $Q$ for simplicity.
This can be handled in the standard way (see, e.g., \cite[Sec. 4.5.3]{Elgammal}).
The encoding scheme at Node 1 and Node 2 multiplexes two descriptions,
which are obtained in two encoding stages. In the first encoding stage,
the distributed source coding strategy of \cite{Tung}, conventionally
referred to as the Berger-Tung scheme, is adopted by Node 1 and Node
2 to convey descriptions $V_{1}^{n}$ and $V_{2}^{n}$, respectively,
to Node 3. In order for the decoder to be able to recover these descriptions
the rates $R_{1}^{'}$ and $R_{2}^{'}$ allocated by Node 1 and Node
2 have to satisfy the conditions \cite{Tung}\cite[Chapter 13]{Elgammal}\begin{subequations}\label{DSC_ach_reg_NC-1}
\begin{align}
R_{1}^{'} & \geq I(X_{1};V_{1}|V_{2})\label{ach_R1_NC-1}\\
R_{2}^{'} & \geq I(X_{2};V_{2}|V_{1})\label{ach_R2_NC-1}\\
\text{and }R_{1}^{'}+R_{2}^{'} & \geq I(X_{1},X_{2};V_{1},V_{2}).\label{ach_R1+2_NC-1}
\end{align}
\end{subequations}Having decoded the descriptions $(V_{1}^{n},V_{2}^{n})$,
Node 3 selects the action sequence $A^{n}$ as the per-symbol function
$A_{i}=\mathrm{a}(V_{1i},V_{2i})$ for $i\in[1,n]$. Node 3 thus measures
the side information sequence $Y^{n}$. The sequences $(Y^{n},V_{1}^{n},V_{2}^{n})$
can then be regarded as side information available at the decoder.
Therefore, in the second encoding stage, the distributed Wyner-Ziv
scheme proposed in \cite[Theorem 2]{Gastpar} is used to convey the
descriptions $U_{1}^{n}$ and $U_{2}^{n}$ by Node 1 and Node 2, respectively,
to Node 3. Note that the fact that sequences $(Y^{n},V_{1}^{n},V_{2}^{n})$
are not i.i.d. does not affect the achievability of the rate region
derived in \cite{Gastpar}. This is because, as shown in \cite[Lemma 3.1]{Elgammal},
the packing lemma leveraged to ensure the correctness of the decoding
process applies for an arbitrary distribution of the sequences $(Y^{n},V_{1}^{n},V_{2}^{n})$.
In order for the decoder to correctly retrieve the descriptions $U_{1}^{n}$
and $U_{2}^{n}$, the rates $R_{1}^{''}$ and $R_{2}^{''}$ allocated
by Node 1 and Node 2 must satisfy the inequalities \cite{Gastpar}
\begin{subequations}\label{DSC_ach_reg_NC-2} 
\begin{align}
R_{1}^{''} & \geq I(X_{1};U_{1}|V_{1},V_{2},U_{2},Y)\label{ach_R1_NC-2}\\
R_{2}^{''} & \geq I(X_{2};U_{2}|V_{1},V_{2},U_{1},Y)\label{ach_R2_NC-2}\\
\text{and }R_{1}^{''}+R_{2}^{''} & \geq I(X_{1},X_{2};U_{1},U_{2}|V_{1},V_{2},Y).\label{ach_R1+2_NC-2}
\end{align}
\end{subequations} Node 1 and Node 2 multiplex the source indices
obtained in the two phases and hence the overall rates are $R_{1}=R_{1}^{'}+R_{1}^{''}$
and $R_{2}=R_{2}^{'}+R_{2}^{''}$. Using these equalities, along with
(\ref{DSC_ach_reg_NC-1}) and (\ref{DSC_ach_reg_NC-2}), leads to
(\ref{DSC_ach_reg_NC}). Finally, the decoder $j$ estimates $\hat{X}_{j}^{n}$
with $j=1,2$ sample by sample as a function of $U_{1i},U_{2i}$ and
$Y_{i}$. The proof of the cardinality bounds follows from standard
arguments and is sketched in Appendix A%
\footnote{It is noted that, using the approach of \cite{Jana}, it may be possible
to improve the cardinality bounds. This aspect is not further explored
here.%
}. We now turn to a similar achievable strategy for the case with causal
side information. 

\begin{proposition}\label{prop:ach_C} The rate-distortion-cost region
with causal side information at Node 3 satisfies the inclusion ${\cal R}_{C}(D_{1},D_{2},\Gamma)\supseteq{\cal R}_{C}^{a}(D_{1},D_{2},\Gamma)$,
where the region ${\cal R}_{C}^{a}(D_{1},D_{2},\Gamma)$ is given
by the union of the set of all of rate tuples $(R_{1},R_{2})$ that
satisfy the inequalities\emph{ }\begin{subequations}\emph{\label{DSC_ach_reg_C}}
\begin{align}
R_{1} & \geq I(X_{1};U_{1}|U_{2},Q)\label{ach_R1_C}\\
R_{2} & \geq I(X_{2};U_{2}|U_{1},Q)\label{ach_R2_C}\\
\text{and }R_{1}+R_{2} & \geq I(X_{1},X_{2};U_{1},U_{2}|Q),\label{ach_R1+R2_C}
\end{align}
\end{subequations} for some joint pmfs that factorizes as 
\begin{eqnarray}
p(q,x_{1},x_{2},y,u_{1},u_{2},a,\hat{x}_{1},\hat{x}_{2}) & = & p(q)p(x_{1},x_{2})p(u_{1}|x_{1},q)p(u_{2}|x_{2},q)\delta(a-\mathrm{a}(u_{1},u_{2},q))\nonumber \\
 &  & p(y|a,x_{1},x_{2})\delta(\hat{x}_{1}-\mathrm{\hat{x}}_{1}(u_{1},u_{2},y,q))\nonumber \\
 &  & \delta(\hat{x}_{2}-\mathrm{\hat{x}}_{2}(u_{1},u_{2},y,q)),\label{eq:joint_ach_C}
\end{eqnarray}
with pmfs $p(q)$, $p(u_{1}|x_{1},q)$ and $p(u_{2}|x_{2},q)$ and
deterministic functions $\mathrm{a}\textrm{\ensuremath{\mathrm{:}} }\mathcal{U}_{1}\times\mathcal{U}_{2}\times{\cal Q}\rightarrow\mathcal{A}$
and $\mathrm{\hat{x}}_{j}\textrm{\ensuremath{\mathrm{:}} }\mathcal{U}_{1}\times\mathcal{U}_{2}\times\mathcal{Y}\times Q\rightarrow\hat{\mathcal{X}}_{j}$
for $j=1,2$, such that the action and the distortion constraints
(\ref{eq:action_const_ach})-(\ref{eq:dist_const_ach}) hold, respectively.
Finally, any extreme point in the region $\mathcal{R}_{C}^{a}(D_{1},D_{2},\Gamma)$
can be obtained by constraining the cardinalities of random variables
$(U_{1},U_{2})$ as $\left\vert \mathcal{U}_{1}\right\vert \leq\left\vert \mathcal{X}_{1}\right\vert +5$
and $\left\vert \mathcal{U}_{2}\right\vert \leq\left\vert \mathcal{X}_{2}\right\vert +5$.
\end{proposition}

The proof follows by similar arguments as the ones in the proof of
Proposition \ref{prop:ach_NC} with the only difference that only
one stage of encoding is sufficient. Specifically, as in Proposition
\ref{prop:ach_NC}, Berger-Tung coding is adopted to convey the descriptions
$U_{1}^{n}$ and $U_{2}^{n}$ to Node 3. Note that, with causal side
information, there is no advantage in having a second encoding stage,
since the side information sequence cannot be leveraged for binning
in contrast to the case with non-causal side information \cite{Elgammal Weissman}\cite[Chapter 12]{Elgammal}.
The cardinality bounds follow from arguments similar to Appendix A.

\subsection{Degraded Source Sets and Causal Side Information\label{sec:XY,Y-Causal}}

In this section, we consider the special case in which the sequence
observed by Node 2 is a symbol-by-symbol function of the source observed
at Node 1 \cite[Sec. V.]{Kaspi-1} (see also \cite{Wagner}). In other
words, we can write $X_{1i}=(X_{1i}^{'},X_{2i})$ for $i\in[1,n]$,
where $X{}_{1}^{'n}$ is an i.i.d. sequence independent of $X_{2}^{n}$.
We refer to this set-up as having \emph{degraded source sets}. Moreover,
we assume that the side information $Y$ is available causally at
Node 3. The next proposition proves that the achievable strategy of
Proposition \ref{prop:ach_C} is optimal in this case.

\begin{proposition}\label{prop:DSC_opt1} The rate-distortion-cost
region $\mathcal{R}_{C}(D_{1},D_{2},\Gamma)$ for the set-up with
degraded source sets and with causal side information at Node 3 satisfies
$\mathcal{R}_{C}(D_{1},D_{2},\Gamma)=\mathcal{R}_{C}^{a}(D_{1},D_{2},\Gamma)$.
\end{proposition}

\begin{remark}Proposition 3 generalizes to the case with action-dependent
side information the result in \cite[Sec. V]{Kaspi-1} for the case
with no side information. \end{remark}

For the proof of converse, we refer the reader to Appendix B.

\subsection{One-Distortion Criterion and Non-Causal Side Information \label{sec:distsc-X,Y-NonCausal-lossy-lossless}}

In this section, we consider a variation on the set-up of source coding
with action-dependent non-causal side information described in Definition
\ref{def1}. Specifically, Node 3 selects the action sequence $A^{n}$
based only on the message $M_{1}$ received from Node 1. In other
words, the action function (\ref{action enc}) is modified to 
\begin{equation}
\mathrm{\ell}\text{: }[1,2^{nR_{1}}]\rightarrow\mathcal{A}^{n},\label{action enc-1}
\end{equation}
which maps the message $M_{1}$ into an action sequence $A^{n}$ at
Node 3. This may be the case in scenarios in which there is a hierarchy
between Node 1 and Node 2, e.g., in a sensor network, and the functionality
of remote control of the side information is assigned solely to Node
1. The next proposition characterizes the rate-distortion-cost function
$\mathcal{R}_{NC}(D_{1},0,\Gamma)$ under the mentioned assumption
when Hamming distortion is selected for $\hat{X}_{2}$. That is, we
choose the distortion measure $d_{2}(x_{2},\hat{x}_{2})$ as $d_{H}(x_{2},\hat{x}_{2})=0$
if $x_{2}=\hat{x}_{2}$ and $d_{H}(x_{2},\hat{x}_{2})=1$ otherwise.
This implies that we impose the constraint of vanishingly small per-symbol
Hamming distortion between source $X_{2}^{n}$ and estimate $\hat{X}_{2}^{n}$,
or equivalently the constraint $\frac{1}{n}\overset{n}{\underset{i=1}{\sum}}\Pr[\hat{X}_{2i}\neq X_{2i}]\rightarrow0$
for $n\rightarrow\infty$. We will refer to this assumption by saying
that source sequence $X_{2}^{n}$ must be recovered losslessly at
the decoder.

\begin{proposition} \label{prop:DSC_opt2}If the action function
is given by (\ref{action enc-1}) and $X_{2}^{n}$ must be recovered
losslessly at Node 3, the \textit{\emph{rate-distortion-cost}}\textit{
}\textit{\emph{region}}\textit{ }$\mathcal{R}_{NC}(D_{1},0,\Gamma)$
is given by union of the set of all of rate tuples $(R_{1},R_{2})$
that satisfy the inequalities \begin{subequations}\label{DSC_opt2}
\begin{align}
R_{1} & \geq I(X_{1};A|Q)+I(X_{1};U_{1}|A,X_{2},Y,Q)\label{R1_opt2}\\
R_{2} & \geq H(X_{2}|A,Y,V,Q)\label{R2_opt2}\\
\text{and }R_{1}+R_{2} & \geq I(X_{1};A|Q)+H(X_{2}|A,Y,Q)+I(X_{1};U_{1}|A,X_{2},Y,Q),\label{R1+R2_opt2}
\end{align}
for some joint pmfs that factorizes as\end{subequations} 
\begin{equation}
p(q,x_{1},x_{2},y,u_{1},a,\hat{x}_{1})=p(q)p(x_{1},x_{2})p(a,u_{1}|x_{1},q)p(y|a,x_{1},x_{2})\delta(\hat{x}_{1}-\mathrm{\hat{x}}_{1}(u_{1},y,q)),\label{eq:joint_opt2}
\end{equation}
with pmfs $p(q)$ and $p(a,u_{1}|x_{1},q)$ and deterministic function
$\hat{\textrm{x}}_{1}(u_{1},y,q)$, such that the action and the distortion
constraints\begin{subequations}\label{DSC_opt2_const}
\begin{eqnarray}
\mathrm{E}\left[\Lambda(A)\right] & \leq & \Gamma\label{eq:action_const_opt2}\\
\text{and }\mathrm{E}\left[d_{1}(X_{1},X_{2},Y,\hat{X}_{1})\right] & \leq & D_{1}\label{eq:dist_const_opt2}
\end{eqnarray}
\end{subequations}hold. Finally, $Q$ and $U_{1}$ are auxiliary
random variables whose alphabet cardinality can be constrained as
$\left\vert Q\right\vert \leq6$ and $\left\vert \mathcal{U}_{1}\right\vert \leq6\left\vert \mathcal{X}_{1}\right\vert \left\vert \mathcal{A}\right\vert +3$
without loss of optimality. \end{proposition}

\begin{remark} In the case in which there is no side information,
Proposition \ref{prop:DSC_opt2} reduces to \cite[Theorem 1]{Berger-Yeung}.
\end{remark}

For the proof of converse, we refer the reader to Appendix C. The
achievability follows from Proposition \ref{prop:ach_NC} by setting
$V_{2}=\emptyset$, $V_{1}=A$ and $U_{2}=X_{2}$.

\begin{remark}Extension of the result in Proposition to an arbitrary
number $K$ of encoders can be found in \cite{Ahmadi_DSC}.\end{remark}

\subsection{A Binary Example\label{example}}

We now focus on a specific numerical example in order to illustrate
the result derived in Proposition \ref{prop:ach_NC} and Proposition
\ref{prop:DSC_opt2} and the advantage of selecting actions at Node
3 based on the message received from one of the nodes. Specifically,
we assume that all alphabets are binary and that ($X_{1},X_{2})$
is a doubly symmetric binary source (DSBS) characterized by probability
$p,$ with $0\leq p\leq1/2$, so that $p(x_{1})=p(x_{2})=1/2$ for
$x_{1},x_{2}\in\{0,1\}$ and $\Pr[X_{1}\neq X_{2}]=p$. Moreover,
we adopt Hamming distortion for both sources to reconstruct both $X_{1}$
and $X_{2}$ losslessly in the sense discussed above. Note that, this
implies that we set $d_{1}(x_{1},x_{2},y,\hat{x}_{1})=d_{H}(x_{1},\hat{x}_{1})$
and $D_{1}=0.$ The side information $Y_{i}$ is such that 
\begin{equation}
Y_{i}=\left\{ \begin{array}{c}
\mathrm{f}(X_{1i},X_{2i})\text{ \ \ \ \ \ \ \ \ \ \ \ \ \ if }A_{i}=1\\
1\text{\ \ \ \ \ \ \ \ \ \ \ \ \ \ \ \ \ \ \ \ \ \ \ \ \ \ if }A_{i}=0
\end{array}\right.,\label{ex}
\end{equation}
where $\mathrm{f}(x_{1},x_{2})$ is a deterministic function to be
specified. Therefore, when action $A_{i}=1$ is selected, then $Y_{i}=\mathrm{f}(X_{1i},X_{2i})$
is measured at the receiver, while\ with $A_{i}=0$ no useful information
is collected by the decoder. The action sequence $A^{n}$ must satisfy
the cost constraint (\ref{action_cost_const}), where the cost function
is defined as $\Lambda(A_{i})=1$ if $A_{i}=1$ and $\Lambda(A_{i})=0$
if $A_{i}=0$. It follows that, given (\ref{ex}), a cost $\Gamma$
implies that the decoder can observe $\mathrm{f}(X_{1i},X_{2i})$
only for at most $n\Gamma$ symbols. As for the function $\mathrm{f}(x_{1},x_{2}),$
we consider two cases, namely $\mathrm{f}(x_{1},x_{2})=x_{1}\oplus x_{2},$
where $\oplus$ is the binary sum and $\mathrm{f}(x_{1},x_{2})=x_{1}\odot x_{2}$,
where $\odot$ is the binary product. We assume that the side information
is available non-causally at the decoder.

\begin{figure}[h!]
\centering\includegraphics[bb=49bp 196bp 547bp 578bp,clip,width=4.0906in,height=2.9334in]{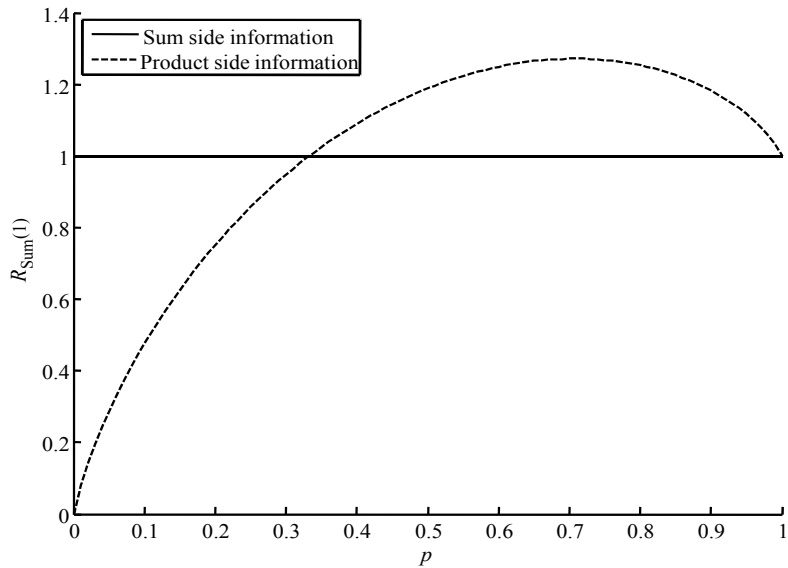}
\caption{Sum-rates versus $p$ for sum and product side informations ($\Gamma=1$).}

\label{fig:plot3} 
\end{figure}

To start with, observe that the sum-rate is a non-increasing function
of the action cost $\Gamma$ and hence the minimum sum-rate is obtained
when $\Gamma=1$. With $\Gamma=1,$ it is clearly optimal to set $A=1,$
irrespective of the value of $X_{1}$. In this case, from the Slepian-Wolf
theorem, the sum rate equals $R_{sum}(1)=H(X_{1},X_{2}|Y)$. Specifically,
with sum side information we get
\begin{equation}
R_{sum}^{\oplus}(1)=1,\label{sumrate_sum1}
\end{equation}
since we have $R_{sum}^{\oplus}(1)=H(X_{1},X_{2}|X_{1}\oplus X_{2})=H(X_{1}|X_{1}\oplus X_{2})=H(X_{1}),$
where the second equality follows from the chain rule and the second
from the crypto-lemma \cite[Lemma 2]{Forney}. Instead, with product
side information, we obtain
\begin{equation}
R_{sum}^{\odot}(1)=H\left(\frac{1-p}{1+p},\frac{p}{1+p},\frac{p}{1+p}\right)\left(\frac{1+p}{2}\right),\label{sumrate_pro1}
\end{equation}
where we have used the definition $H\left(p_{1},p_{2},...,p_{k}\right)=-\sum_{i=1}^{k}p_{k}\log_{2}p_{k}.$
Equation (\ref{sumrate_pro1}) follows since
\begin{align}
R_{sum}^{\odot}(1) & =H(X_{1},X_{2}|X_{1}\odot X_{2})\nonumber \\
 & =H(X_{1},X_{2}|X_{1}\odot X_{2}=0)\Pr[X_{1}\odot X_{2}=0],\label{sumrate_pro21}
\end{align}
where the second equality is a consequence of the fact that $X_{1}\odot X_{2}=1$
implies that $X_{1}=1$ and $X_{2}=1.$ Sum-rate (\ref{sumrate_pro1})
is then obtained by evaluating (\ref{sumrate_pro21}) for the DSBS
at hand. Fig. \ref{fig:plot3}
shows the sum-rates (\ref{sumrate_sum1}) and (\ref{sumrate_pro1}),
demonstrating that, if $p$ is sufficiently small, namely if $p\lesssim0.33,$
we have $R_{sum}^{\odot}(1)<R_{sum}^{\oplus}(1)$ and thus product
side information is more informative than the sum, while for $p\gtrsim0.33$
the opposite is true (and for $p=1,$ they are equally informative).
\begin{figure}[h!]
\centering\includegraphics[width=4.2973in,height=3.096in]{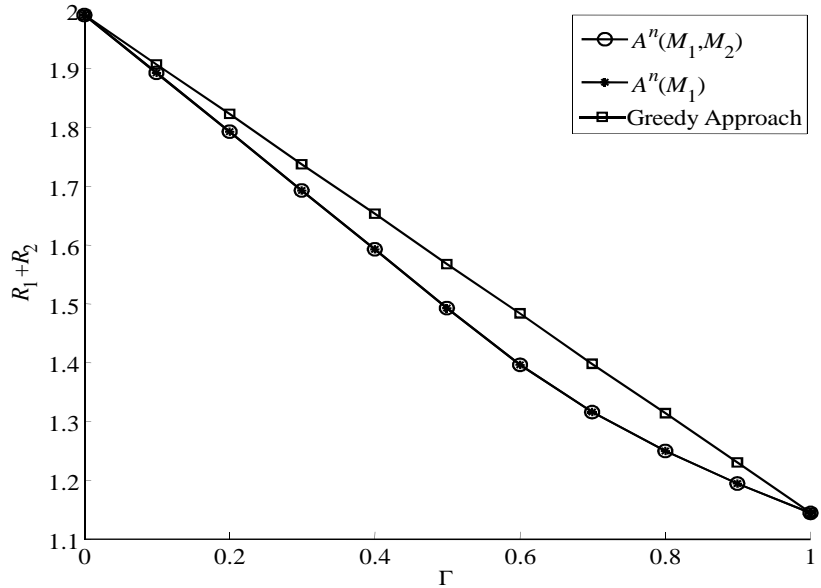}
\caption{Sum-rates versus the action cost $\Gamma$ for product side information
($p=0.45$).}

\label{fig:plot4_new} 
\end{figure}
\begin{figure}[h!]
\centering\includegraphics[width=4.2973in,height=3.096in]{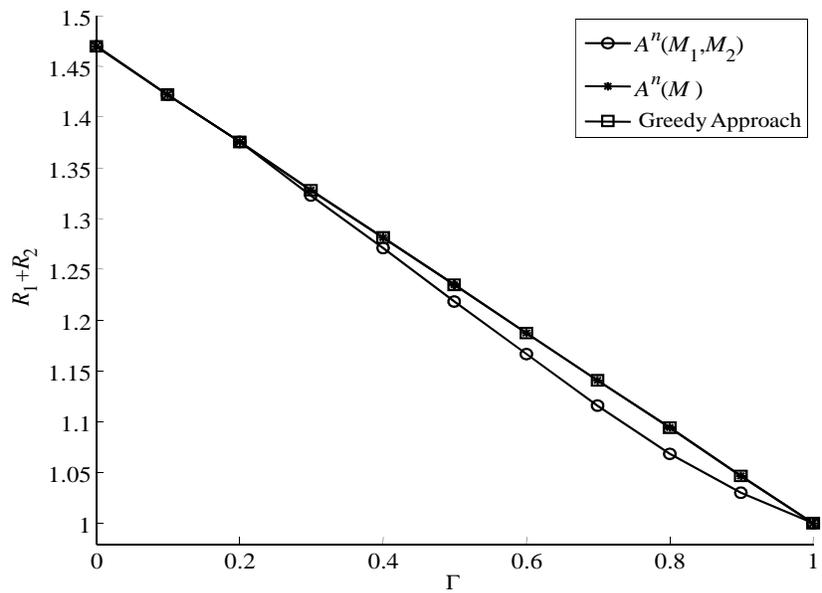}
\caption{Sum-rates versus the action cost $\Gamma$ for sum side information
($p=0.1$).}

\label{fig:plot5_new} 
\end{figure}

Considering a general cost budget $0\leq\Gamma\leq1$, in order to
emphasize the role of both data and control information for the system
performance, we now evaluate the sum-rate attainable by imposing that
the action $A$ be selected by Node 3 a priori, that is, without any
control from Node 1. This can be easily seen to be given by \cite{Permuter}
\begin{align}
R_{sum,\text{ }greedy}(\Gamma) & =\Gamma H(X_{1},X_{2}|Y)+(1-\Gamma)H(X_{1},X_{2})\nonumber \\
 & =\Gamma H(X_{1},X_{2}|Y)+(1-\Gamma)(1+H(p)).\label{Rsumgreedy}
\end{align}
This sum-rate will be compared below with the performance of the scheme
in Proposition \ref{prop:ach_NC}, in which the actions are selected
based on both messages $(M_{1},M_{2})$, and that of Proposition \ref{prop:DSC_opt2},
in which the actions are selected based only on message $M_{1}$. 

Fig. \ref{fig:plot4_new} depicts the mentioned sum-rates%
\footnote{The sum-rate from Proposition \ref{prop:ach_NC} is calculated by
assuming binary auxiliary variables $V_{1}$ and $V_{2}$ and performing
global optimization.%
} versus the action cost $\Gamma$ for $p=0.45$ and product side information.
It can be seen that the greedy approach suffers from a significant
performance loss with respect to the approaches in which actions are
selected based on the messages received from one encoder or both encoders.
It can be also observed that no gains are obtained by selecting the
actions based on both messages. The fact that choosing the action
based on the message received from Node 1 provides performance benefits
can be explained as follows. If $X_{1}=0,$ the value of the side
information is always $Y=X_{1}\odot X_{2}=0$ irrespective of the
value of $X_{2}.$ Therefore, if $X_{1}=0,$ the side information
is less informative than if $X_{1}=1$ and hence it may be advantageous
to save on the action cost by setting $A=0.$ Consequently, choosing
actions based on the message received from Node 1 can result in a
lower sum-rate.

The scenario with sum side information is considered in Fig. \ref{fig:plot5_new}
for $p=0.1$. A first observation is that, as proved in Appendix D,
choosing the action based only on $M_{1}$ cannot improve the sum-rate
with respect to the greedy case. This contrasts with the product side
information case, and is due to the fact that $X_{1}$ is independent
of the side information $Y$. Instead, choosing the actions based
on both messages allows to save on the necessary communication sum-rate.

\section{Cascade Source Coding with a Side Information Vending Machine\label{sec:cascadesc}}

In this section, we first describe the system model for the setting
of Fig. \ref{fig:fig3} of cascade source coding with a side information
vending machine. We recall that side information $Y$\ is here assumed
to be available causally at the decoder (Node 3). The corresponding
model with non-causal side information is studied in \cite{Ahmadi_Chiru}.
We then present the characterization of the corresponding rate-distortion-cost
performance in Sec. \ref{rate_dist_cost_cascade}.

\subsection{System Model}

The problem of cascade lossy computing with causal observation costs
at second user, illustrated in Fig. \ref{fig:fig3}, is defined by
the pmfs $p_{X_{1}X_{2}}(x_{1},x_{2})$ and $p_{Y|AX_{1}X_{2}}(y|a,x_{1},x_{2})$
and discrete alphabets $\mathcal{X}_{1},\mathcal{X}_{2},\mathcal{Y},\mathcal{A},\mathcal{\hat{X}}_{1},\mathcal{\hat{X}}_{2},$
as follows. The source sequences $X_{1}^{n}$ and $X_{2}^{n}$ with
$X_{1}^{n}\in\mathcal{X}_{1}^{n}$ and $X_{2}^{n}\in\mathcal{X}_{2}^{n}$,
respectively$,$ are such that the pairs $(X_{1i},X_{2i})$ for $i\in[1,n]$
are i.i.d. with joint pmf $p_{X_{1}X_{2}}(x_{1},x_{2})$. Node 1 measures
sequences $X_{1}^{n}$ and $X_{2}^{n}$ and encodes them in a message
$M_{12}$ of $nR_{12}$ bits, which is delivered to Node 2. Node 2
estimates a sequence $\hat{X}_{1}^{n}\in\mathcal{\hat{X}}_{1}^{n}$
within given distortion requirements to be discussed below. Moreover,
Node 2 encodes the message $M_{12}$, received from Node 1, and the
locally available sequence $X_{2}^{n}$ in a message $M_{23}$ of
$nR_{23}$ bits, which is delivered to node 3. Node 3 wishes to estimate
a sequence $\hat{X}_{2}^{n}\in\mathcal{\hat{X}}_{2}^{n}$ within given
distortion requirements to be discussed. To this end, Node 3 receives
message $M_{23}$ and based on this, selects an action sequence $A^{n},$
where $A^{n}\in\mathcal{A}^{n}.$ The action sequence affects the
quality of the measurement $Y^{n}$ of sequence $X_{1}^{n}$ and $X_{2}^{n}$
obtained at the Node 3. Specifically, given $A^{n}$, $X_{1}^{n}$
and $X_{2}^{n}$, the sequence $Y^{n}$ is distributed as in (\ref{pzxya}).
The cost of the action sequence is defined by a cost function $\Lambda$:
$\mathcal{A\rightarrow}[0,\Lambda_{\max}]$ with $0\leq\Lambda_{\max}<\infty,$
as in (\ref{overall_cost}). The estimated sequence $\hat{X}_{2}^{n}$
with $\hat{X}_{2}^{n}\in\mathcal{\hat{X}}_{2}^{n}$ is then obtained
as a function of $M_{23}$ and $Y^{n}$.

Estimated sequences $\hat{X}_{j}^{n}$ for $j=1,2$ must satisfy distortion
constraints defined by functions $d_{j}(x_{1},x_{2},y,\hat{x}_{j})$:
$\mathcal{X}_{1}\times\mathcal{X}_{2}\times\mathcal{Y}\times\mathcal{\hat{X}}_{j}\rightarrow\lbrack0,D_{\max}]$
with $0\leq D_{\max}<\infty$ for $j=1,2,$ respectively. A formal
description of the operations at encoder and decoder follows.

\begin{definition} \label{def1_cascade}An $(n,R_{12},R_{23},D_{1},D_{2},\Gamma)$
code for the set-up of Fig. \ref{fig:fig3} consists of two source
encoders, namely
\begin{equation}
\mathrm{g}_{1}\text{: }\mathcal{X}_{1}^{n}\times\mathcal{X}_{2}^{n}\rightarrow[1,2^{nR_{12}}],\label{encoder1}
\end{equation}
which maps the sequences $X_{1}^{n}$ and $X_{2}^{n}$ into a message
$M_{12};$
\begin{equation}
\mathrm{g}_{2}\text{:}\text{ }\mathcal{X}_{2}^{n}\times[1,2^{nR_{12}}]\rightarrow[1,2^{nR_{23}}]\label{encoder2}
\end{equation}
which maps the sequence $X_{2}^{n}$ and message $M_{12}$ into a
message $M_{23};$ an ``action\textquotedblright{}\ function 
\begin{equation}
\mathrm{\ell}\text{: }[1,2^{nR_{23}}]\rightarrow\mathcal{A}^{n},\label{action_fun}
\end{equation}
which maps the message $M_{23}$ into an action sequence $A^{n};$
a decoding function
\begin{equation}
\mathrm{h}_{1}\text{: }[1,2^{nR_{12}}]\times\mathcal{X}_{2}^{n}\rightarrow\mathcal{\hat{X}}_{1}^{n},\label{decoder1}
\end{equation}
which maps the message $M_{12}$ and the measured sequence $X_{2}^{n}$
into the estimated sequence $\hat{X}_{1}^{n};$ and a sequence of
decoding functions
\begin{equation}
\mathrm{h}_{2i}\text{: }[1,2^{nR_{23}}]\times\mathcal{Y}^{i}\rightarrow\mathcal{\hat{X}}_{2},\label{decoder2}
\end{equation}
for $i\in[1,n]$ which maps the message $M_{23}$ and the measured
sequence $Y^{i}$ into the $i$th estimated symbol $\hat{X}_{2i}=\mathrm{h}_{2i}(M_{23},Y^{i});$
such that the action cost constraint $\Gamma$ and distortion constraints
$D_{j}$ for $j=1,2$ are satisfied, i.e.,
\begin{align}
\frac{1}{n}\underset{i=1}{\overset{n}{\sum}}\mathrm{E}\left[\Lambda(A_{i})\right] & \leq\Gamma\label{action cost}\\
\text{ and }\frac{1}{n}\underset{i=1}{\overset{n}{\sum}}\mathrm{E}\left[d_{j}(X_{1i},X_{2i},Y_{i},\hat{X}_{ji})\right] & \leq D_{j}\text{ for }j=1,2,\label{dist const}
\end{align}
respectively. \end{definition}

\begin{definition} \label{def2_cascade}Given a distortion-cost tuple
$(D_{1},D_{2},\Gamma)$, a rate tuple $(R_{12},R_{23})$ is said to
be achievable if, for any $\epsilon>0$, and sufficiently large $n$,
there exists a $(n,R_{12},R_{23},D_{1}+\epsilon,D_{2}+\epsilon,\Gamma+\epsilon)$
code.\end{definition}

\begin{definition} The \textit{rate-distortion-cost region }$\mathcal{R}(D_{1},D_{2},\Gamma)$
is defined as the closure of all rate tuples $(R_{12},R_{23})$ that
are achievable given the distortion-cost tuple $(D_{1},D_{2},\Gamma)$.
\end{definition} 

\begin{remark} For side information $Y$ independent of the action
$A$ given $X_{1}$ and $X_{2},$ i.e., for $\ p(y|a,x_{1},x_{2})=p(y|x_{1},x_{2}),$
the rate-distortion region $\mathcal{R}(D_{1},D_{2},\Gamma)$ has
been derived in \cite{Chia-causal}. \end{remark}

\subsection{Rate-Distortion-Cost Region\label{rate_dist_cost_cascade}}

We have the following characterization of the rate-distortion-cost
region.

\begin{proposition} \label{prop:3} The rate-distortion-cost region
$\mathcal{R}(D_{1},D_{2},\Gamma)$ for the set-up of Fig. \ref{fig:fig3}
is given by the union of all rate pairs ($R_{12},R_{23}$) satisfying
the inequalities \begin{subequations} \label{reg_cascade}
\begin{align}
R_{12} & \geq I(X_{1};U,A,\hat{X}_{1}|X_{2})\label{reg1_1}\\
\text{and }R_{23} & \geq I(X_{1},X_{2};U,A),\label{reg1}
\end{align}
for some joint pmf that factorizes as\end{subequations} 
\begin{align}
p(x_{1},x_{2},y,a,u,\hat{x}_{1},\hat{x}_{2}) & =p(x_{1},x_{2})p(a,u,\hat{x}_{1}|x_{1},x_{2})p(y|a,x_{1},x_{2})\nonumber \\
 & \cdot\delta(\hat{x}_{2}-\mathrm{\hat{x}}_{2}(u,y)),\label{joint1}
\end{align}
with pmf $p(a,u,\hat{x}_{1}|x_{1},x_{2})$ and deterministic function
$\mathrm{\hat{x}}_{2}(u,y)$, such that the action and the distortion
constraints
\begin{align}
\mathrm{E}\left[\Lambda(A)\right] & \leq\Gamma\label{action_cascade}\\
\text{and }\mathrm{E}[d_{j}(X_{1},X_{2},Y,\hat{X}_{j})] & \leq D_{j},\text{for }j=1,2,\label{dist_cascade}
\end{align}
respectively, hold. Finally, $U$ is an auxiliary random variable
whose alphabet cardinality can be constrained as $|\mathcal{U}|\leq\left\vert \mathcal{X}_{1}\right\vert \left\vert \mathcal{X}_{2}\right\vert +4$,
without loss of optimality. \end{proposition}

\begin{remark} If $p(y|a,x_{1},x_{2})=p(y|x_{1},x_{2}),$ Proposition
\ref{prop:3} reduces to \cite[Theorem 1]{Chia-causal}. \end{remark}

The proof of converse is provided in Appendix E. The coding strategy
that proves achievability is a combination of the techniques proposed
in \cite{Permuter} and \cite[Theorem 1]{Chia-causal}. Here we briefly
outline the main ideas, since the technical details follow from standard
arguments. In the scheme at hand, Node 1 first maps sequences $X_{1}^{n}$
and $X_{2}^{n}$ into the action sequence $A^{n}$ and an auxiliary
codeword $U^{n}$ using the standard joint typicality criterion. This
mapping operation requires a codebook of rate $I(X_{1},X_{2};U,A)$
(see, e.g., \cite[Chapter 3]{Elgammal})$.$ Then, given the so obtained
sequences $A^{n}$ and $U^{n},$ source sequences $X_{1}^{n}$ and
$X_{2}^{n}$ are further mapped into the estimate $\hat{X}_{1}^{n}$
for Node 2 so that the sequences $(X_{1}^{n},X_{2}^{n},A^{n},U^{n},\hat{X}_{1}^{n})$
are jointly typical. This requires rate $I(X_{1},X_{2};\hat{X}_{1}|U,A)$
\cite[Chapter 3]{Elgammal}$.$ Leveraging the side information $X_{2}^{n}$
available at Node 2, conveying the codewords $A^{n},$ $\hat{X}_{1}^{n}$
and $U^{n}$ to Node 2 requires rate $I(X_{1},X_{2};U,A)+I(X_{1},X_{2};\hat{X}_{1}|U,A)-I(U,A,\hat{X}_{1};X_{2})$
\cite[Chapter 12]{Elgammal}$,$ which equals the right-hand side
of (\ref{reg1_1}). Node 2 conveys $U^{n}$ and $A^{n}$ to Node 3
by simply forwarding the index received from Node 1\ (of rate $I(X_{1},X_{2};U,A)$).
Finally, Node 3 estimates $\hat{X}_{2}^{n}$ through a symbol-by-symbol
function as $\hat{X}_{2i}=\mathrm{\hat{x}}_{2}(U_{i},Y_{i})$ for
$i\in[1,n].$

\section{Concluding Remarks\label{remarks}}

In the setting of source coding with a side information vending machine
introduced in \cite{Permuter}, the decoder can control the quality
of the side information through a control, or action, sequence that
is selected based on the message encoded by the source node. Since
this message must also carry information directly related to the source
to be reproduced at the decoder, a key aspect of the model is the
interplay between encoding data and control information.

In this work, we have generalized the original work \cite{Permuter}
to two standard multiterminal scenarios, namely distributed source
coding and cascade source coding. For the former, we obtained inner
bounds to the rate-distortion-cost regions for the cases with non-causal
and causal side information at the decoder. These bounds have been
found to be tight in two special cases. We have also provided some
numerical example to shed some light on the advantages of an optimized
trade-off between data and control transmission. As for the cascade
source coding problem, a single-letter characterizations of achievable
rate-distortion-cost trade-offs has been derived under the assumption
of causal side information at the decoder.

A number of open problems have been left unsolved by this work, including
the identification of more general conditions under which the inner
bounds of Proposition 1 and Proposition 2 are tight. The technical
challenges that we have faced in this task are related to the well-known
issues that arise when identifying auxiliary random variables that
satisfy the desired Markov chain conditions in distributed source
coding problems (see, e.g., \cite[Chapter 13]{Elgammal}).

\appendices{}

\section*{Appendix A}

Using standard inequalities, it can be seen that the rate region (\ref{DSC_ach_reg_NC})
evaluated with a constant $Q$ is a contra-polymatroid, as the Berger-Tung
region (\ref{DSC_ach_reg_C}) (see e.g., \cite{Chen-Berger}). Moreover,
the role of the variable $Q$ is that of performing the convexification
of the union of all regions of tuples $(R_{1},R_{2},D_{1},D_{2},\Gamma)$
that satisfy (\ref{DSC_ach_reg_NC}) and (\ref{DSC_General_const})
for some fixed $Q$. It follows from \cite{Chen-Berger} that every
extreme point of region of achievable tuples $(R_{1},R_{2},D_{1},D_{2},\Gamma)$
satisfies the equations\begin{subequations} \label{vertex1}
\begin{eqnarray}
R_{1} & = & I(X_{1};V_{1}|V_{2})+I(X_{1};U_{1}|U_{2},V_{1},V_{2},Y)\label{eq:vertex1_R1}\\
R_{2} & = & I(X_{2};V_{2})+I(X_{2};U_{2}|V_{1},V_{2},Y)\label{eq:vertex1_R2}
\end{eqnarray}
\end{subequations} along with (\ref{DSC_General_const}), where both
relationships are satisfied with equality, or \begin{subequations}
\label{vertex2}
\begin{eqnarray}
R_{1} & = & I(X_{1};V_{1})+I(X_{1};U_{1}|V_{1},V_{2},Y)\label{eq:vertex2_R1}\\
R_{2} & = & I(X_{2};V_{2}|V_{1})+I(X_{2};U_{2}|U_{1},V_{1},V_{2},Y)\label{eq:vertex2_R2}
\end{eqnarray}
\end{subequations} along with (\ref{DSC_General_const}) satisfied
with equality. Applying the Fenchel--Eggleston--Caratheodory theorem
to the right-hand side of the equations above and to (\ref{DSC_General_const})
concludes the proof (See \cite[Appendix C]{Elgammal} and \cite{Tung}).

\section*{Appendix B}

\section*{Proof of the Converse for Proposition \ref{prop:DSC_opt1}}

In this section, the proof of converse for Proposition \ref{prop:DSC_opt1}
is given. For any $(n,R_{1},R_{2},D_{1}+\epsilon,D_{2}+\epsilon,\Gamma+\epsilon)$
code, we have the following inequalities:

\begin{align*}
nR_{1} & \geq H(M_{1})\geq H(M_{1}|M_{2})\\
 & \overset{(a)}{=}I(M_{1};X_{1}^{n},X_{2}^{n}|M_{2})\\
 & \overset{}{=}\overset{n}{\underset{i=1}{\sum}}H(X_{1i},X_{2i}|X_{1}^{i-1},X_{2}^{i-1},M_{2})-H(X_{1i},X_{2i}|X_{1}^{i-1},X_{2}^{i-1},M_{1},M_{2})\\
 & \overset{(b)}{=}\overset{n}{\underset{i=1}{\sum}}H(X_{1i},X_{2i}|X_{1}^{i-1},X_{2}^{i-1},M_{2})-H(X_{1i},X_{2i}|X_{1}^{i-1},X_{2}^{i-1},M_{1},M_{2},Y^{i-1})\\
 & \overset{(c)}{\geq}\overset{n}{\underset{i=1}{\sum}}H(X_{1i},X_{2i}|X_{1}^{i-1},X_{2}^{i-1},M_{2},Y^{i-1})-H(X_{1i},X_{2i}|X_{1}^{i-1},X_{2}^{i-1},M_{1},M_{2},Y^{i-1})\\
 & \overset{(d)}{=}\overset{n}{\underset{i=1}{\sum}}I(X_{1i},X_{2i};U_{1i}|U_{2i}),
\end{align*}
where (\textit{a}) follows because $M_{1}$ is a function of $(X_{1}^{n},X_{2}^{n})$
given that $X_{2}^{n}$ is a function of $X_{1}^{n}$ by assumption;
($b$) follows since $(X_{1i},X_{2i})\textrm{---}(X_{1}^{i-1},X_{2}^{i-1},M_{1},M_{2})\textrm{---}Y^{i-1}$forms
a Markov chain; ($c$) follows by the fact that conditioning decreases
entropy; and $(d)$ follows by defining $U_{ji}=(X_{1}^{i-1},X_{2}^{i-1},Y^{i-1},M_{j})$
for $j=1,2$. We also have a similar chain of inequalities for $R_{2}$.
As for the sum-rate $R_{1}+R_{2}$, we have
\begin{align*}
n(R_{1}+R_{2}) & \geq H(M_{1},M_{2})\\
 & \overset{(a)}{=}I(M_{1},M_{2};X_{1}^{n},X_{2}^{n})\\
 & \overset{}{=}\overset{n}{\underset{i=1}{\sum}}H(X_{1i},X_{2i}|X_{1}^{i-1},X_{2}^{i-1})-H(X_{1i},X_{2i}|X_{1}^{i-1},X_{2}^{i-1},M_{1},M_{2})\\
 & \overset{(b)}{=}\overset{n}{\underset{i=1}{\sum}}H(X_{1i},X_{2i}|X_{1}^{i-1},X_{2}^{i-1})-H(X_{1i},X_{2i}|X_{1}^{i-1},X_{2}^{i-1},M_{1},M_{2},Y^{i-1})\\
 & \overset{(c)}{\geq}\overset{n}{\underset{i=1}{\sum}}I(X_{1i},X_{2i};U_{1i},U_{2i}),
\end{align*}
where (\textit{a}) follows because $(M_{1},M_{2})$ are functions
of $(X_{1}^{n},X_{2}^{n})$; ($b$) follows since $(X_{1i},X_{2i})\textrm{---}$
$(X_{1}^{i-1},X_{2}^{i-1},M_{1},M_{2})\textrm{---}Y^{i-1}$ forms
a Markov chain; and ($c$) follows using the definition of $U_{ji}$
for $j=1,2$. Next, let $Q$ be a uniform random variable over the
interval $[1,n]$ and independent of $(X_{1}^{n},X_{2}^{n},U_{1}^{n},U_{2}^{n},Y^{n})$
and define $U_{j}\overset{\Delta}{=}(Q,U_{jQ})$, for $j=1,2$, $X_{1}\overset{\Delta}{=}X_{1Q}$,
$X_{2}\overset{\Delta}{=}X_{2Q}$, $Y\overset{\Delta}{=}Y_{Q}$. Note
that $\hat{X}_{j}$ is a function of $U_{1},U_{2}$ and $Y$ for $j=1,2$.
Moreover, from (\ref{action_cost_const}) and (\ref{eq:dist_const}),
we have
\begin{align}
\Gamma+\epsilon & \geq\frac{1}{n}\underset{i=1}{\overset{n}{\sum}}\mathrm{E}\left[\Lambda(A_{i})\right]=\mathrm{E}[\Lambda(A)]\\
\text{and }D_{j}+\epsilon & \geq\frac{1}{n}\underset{i=1}{\overset{n}{\sum}}\mathrm{E}\left[d_{j}(X_{1i},X_{2i},Y_{i}.\hat{X}_{ji},)\right]=\mathrm{E}[d_{1}(X_{1},X_{2},Y,\hat{X}_{j})],\mbox{ for }j=1,2.
\end{align}

\section*{Appendix C}

\section*{Proof of the Converse for Proposition \ref{prop:DSC_opt2}}

In this section, the proof of converse for Proposition \ref{prop:DSC_opt2}
is given. Fix a code $(n,R_{1},R_{2},D_{1}+\epsilon,\epsilon,\Gamma)$
for an $\epsilon>0$, whose existence for all sufficiently large $n$
is required by the definition of achievability. 

From the distortion constraint for $\hat{X}_{2}$, we have the inequality
\begin{equation}
\epsilon\geq\frac{1}{n}\overset{n}{\underset{i=1}{\sum}}\mathrm{E}[d_{H}(X_{2i},\hat{X}_{2i})]\overset{(a)}{=}\frac{1}{n}\overset{n}{\underset{i=1}{\sum}}p_{e,2i},\label{error}
\end{equation}
where we have defined $p_{e,2i}=\mathrm{Pr}[X_{2i}\neq\hat{X}_{2i}],$
and (\textit{a}) follows from the definition of the metric $d_{H}(x,\hat{x})$
as the Hamming distortion$.$ Moreover, we also have the following
chain of inequalities 
\begin{align}
 & H(X_{2}^{n}|\hat{X}_{2}^{n})\overset{(a)}{\leq}\overset{n}{\underset{i=1}{\sum}}H(X_{2i}|\hat{X}_{2i})\overset{(b)}{\leq}\overset{n}{\underset{i=1}{\sum}}H(p_{e,i})+p_{e,i}\log\left\vert \mathcal{\hat{X}}_{2i}\right\vert \nonumber \\
 & \overset{(c)}{\leq}nH\left(\frac{1}{n}\overset{n}{\underset{i=1}{\sum}}p_{e,i}\right)+n\left(\frac{1}{n}\overset{n}{\underset{i=1}{\sum}}p_{e,i}\right)\log\left\vert \mathcal{\hat{X}}_{2i}\right\vert \nonumber \\
 & \overset{(d)}{\leq}nH(\epsilon)+n\epsilon\log\left\vert \mathcal{\hat{X}}_{ji}\right\vert \nonumber \\
 & \overset{\Delta}{=}n\delta(\epsilon),\label{error1}
\end{align}
where (\textit{a}) follows by conditioning reduces entropy; (\textit{b})
follows by Fano's inequality; (\textit{c}) follows by Jensen's inequality;
and (\textit{d}) follows by (\ref{error}), where $\delta(\epsilon)\rightarrow0$
as $\epsilon\rightarrow0.$ Note that, in the following, we use the
convention in \cite[Chapter 3]{Elgammal} of defining as $\delta(\epsilon)$
any function such that $\delta(\epsilon)\rightarrow0$ as $\epsilon\rightarrow0.$

For rate $R_{1}$, we then have the following series of inequalities
\begin{align}
nR_{1} & \geq H(M_{1})\overset{(a)}{=}H(M_{1},A^{n})\nonumber \\
 & \overset{}{=}H(A^{n})+H(M_{1}|A^{n})\nonumber \\
 & \overset{(b)}{\geq}H(A^{n})-H(A^{n}|X_{1}^{n},X_{2}^{n})+H(M_{1}|A^{n},Y^{n},X_{2}^{n})-H(M_{1}|A^{n},Y^{n},X_{1}^{n},X_{2}^{n})\nonumber \\
 & \overset{}{=}I(A^{n};X_{1}^{n},X_{2}^{n})+I(M_{1};X_{1}^{n}|A^{n},Y^{n},X_{2}^{n})\nonumber \\
 & \overset{}{=}I(A^{n};X_{1}^{n},X_{2}^{n})+H(X_{1}^{n}|A^{n},Y^{n},X_{2}^{n})-H(X_{1}^{n}|A^{n},Y^{n},X_{2}^{n},M_{1})\nonumber \\
 & \overset{}{=}H(X_{1}^{n},X_{2}^{n})-H(X_{1}^{n},X_{2}^{n}|A^{n})+H(X_{1}^{n},X_{2}^{n},Y^{n}|A^{n})-H(Y^{n},X_{2}^{n}|A^{n})\nonumber \\
 & -H(X_{1}^{n}|A^{n},Y^{n},X_{2}^{n},M_{1})\nonumber \\
 & \overset{}{=}H(X_{1}^{n},X_{2}^{n})+H(Y^{n}|A^{n},X_{1}^{n},X_{2}^{n})-H(Y^{n},X_{2}^{n}|A^{n})\nonumber \\
 & -H(X_{1}^{n}|A^{n},Y^{n},X_{2}^{n},M_{1}),\label{eq3}
\end{align}
where (\textit{a}) follows because $A^{n}$ is a function of $M_{1}$
and (\emph{b}) follows because entropy is non-negative and conditioning
decreases entropy. For the first three terms in (\ref{eq3}) we have
\begin{align}
 & H(X_{1}^{n},X_{2}^{n})+H(Y^{n}|A^{n},X_{1}^{n},X_{2}^{n})-H(Y^{n},X_{2}^{n}|A^{n})\nonumber \\
 & \overset{}{=}H(X_{1}^{n},X_{2}^{n})+H(Y^{n}|A^{n},X_{1}^{n},X_{2}^{n})-H(Y^{n}|A^{n})-H(X_{2}^{n}|A^{n},Y^{n})\nonumber \\
 & \overset{(a)}{=}\overset{n}{\underset{i=1}{\sum}}H(X_{1i},X_{2i})+H(Y_{i}|Y^{i-1},A^{n},X_{1}^{n},X_{2}^{n})-H(Y_{i}|Y^{i-1},A^{n})-H(X_{2i}|X_{2}^{i-1},A^{n},Y^{n})\nonumber \\
 & \overset{(b)}{\geq}\overset{n}{\underset{i=1}{\sum}}H(X_{1i},X_{2i})+H(Y_{i}|A_{i},X_{1i},X_{2i})-H(Y_{i}|A_{i})-H(X_{2i}|A_{i},Y_{i})\nonumber \\
 & \overset{}{=}\overset{n}{\underset{i=1}{\sum}}H(X_{1i},X_{2i})-I(Y_{i};X_{1i},X_{2i}|A_{i})-H(X_{2i}|A_{i},Y_{i})\nonumber \\
 & \overset{}{=}\overset{n}{\underset{i=1}{\sum}}H(X_{1i},X_{2i})-H(X_{1i},X_{2i}|A_{i})+H(X_{1i},X_{2i}|A_{i},Y_{i})-H(X_{2i}|A_{i},Y_{i})\nonumber \\
 & \overset{}{=}\overset{n}{\underset{i=1}{\sum}}I(X_{1i},X_{2i};A_{i})+H(X_{1i}|A_{i},Y_{i},X_{2i}),\label{eq4}
\end{align}
where (\textit{a}) follows by the chain rule for entropy and the fact
that $X_{1}^{n},X_{2}^{n}$ are i.i.d. and (\textit{b}) follows since
$Y_{i}$---$(A_{i},X_{1i},X_{2i})$---$(Y^{i-1},A^{n\backslash i},X_{1}^{n\backslash i},X_{2}^{n\backslash i})$
forms a Markov chain, by the definition of problem, and since conditioning
reduces entropy.

Combining (\ref{eq3}) and (\ref{eq4}), and defining $U_{1i}=(A^{n\backslash i},Y^{n\backslash i},X_{2}^{n\backslash i},M_{1}),$
we obtain
\begin{align}
 & nR_{1}\overset{(a)}{\geq}\overset{n}{\underset{i=1}{\sum}}I(X_{1i},X_{2i};A_{i})+H(X_{1i}|A_{i},Y_{i},X_{2i})\nonumber \\
 & -H(X_{1i}|X_{1}^{i-1},A^{n},Y^{n},X_{2}^{n},M_{1})\nonumber \\
 & \overset{(b)}{\geq}\overset{n}{\underset{i=1}{\sum}}I(X_{1i};A_{i})+H(X_{1i}|A_{i},Y_{i},X_{2i})-H(X_{1i}|A^{n},Y^{n},X_{2}^{n},M_{1})\nonumber \\
 & \overset{(c)}{=}\overset{n}{\underset{i=1}{\sum}}(X_{1i};A_{i})+I(X_{1i};U_{1i}|A_{i},Y_{i},X_{2i}),\label{R1_conv_opt2}
\end{align}
where (\textit{a}) follows by the chain rule for entropy; (\textit{b})
follows because mutual information is non-negative and due to the
fact that conditioning decreases entropy; and (\textit{c}) follows
by the definition of mutual information and definition of $U_{1i}$.

Next, we consider the rate $R_{2}.$ We have
\begin{align}
nR_{2} & \geq H(M_{2})\overset{}{\geq}H(M_{2}|A^{n},Y^{n},M_{1})-H(M_{2}|A^{n},Y^{n},M_{1},X_{2}^{n})\nonumber \\
 & \overset{}{=}I(M_{2};X_{2}^{n}|A^{n},Y^{n},M_{1})\nonumber \\
 & \overset{}{=}H(X_{2}^{n}|A^{n},Y^{n},M_{1})-H(X_{2}^{n}|A^{n},Y^{n},M_{1},M_{2})\nonumber \\
 & \overset{(a)}{\geq}H(X_{2}^{n}|A^{n},Y^{n},M_{1})-n\delta(\epsilon)\nonumber \\
 & \overset{}{=}\overset{n}{\underset{i=1}{\sum}}H(X_{2i}|X_{2}^{i-1},A^{n},Y^{n},M_{1})-n\delta(\epsilon)\nonumber \\
 & \overset{(b)}{\geq}\overset{n}{\underset{i=1}{\sum}}H(X_{2i}|A_{i},Y_{i},U_{1i})-n\delta(\epsilon),\label{R2_conv_opt2}
\end{align}
where (\emph{a}) follows because from (\ref{error1}), $H(X_{2}^{n}|A^{n},Y^{n},M_{1},M_{2})\leq H(X_{2}^{n}|\hat{X}_{2}^{n})\leq n\delta(\epsilon)$,
given that $\hat{X}_{2}^{n}$ is a function of $M_{1},$ $M_{2}$
and $Y^{n}$and ($b$) follows using the definition of $U_{1i}$ and
due to the fact that conditioning decreases entropy. For the sum-rate
$R_{1}+R_{2}$, we also have the following series of inequalities
\begin{align}
n(R_{1}+R_{2}) & \geq H(M_{1},M_{2})\overset{(a)}{=}H(M_{1},M_{2},A^{n})\nonumber \\
 & \overset{}{=}H(A^{n})+H(M_{1},M_{2}|A^{n})\nonumber \\
 & \overset{}{\geq}H(A^{n})-H(A^{n}|X_{1}^{n},X_{2}^{n})+H(M_{1},M_{2}|A^{n},Y^{n})\nonumber \\
 & -H(M_{1},M_{2}|A^{n},Y^{n},X_{1}^{n},X_{2}^{n})\nonumber \\
 & \overset{}{=}I(A^{n};X_{1}^{n},X_{2}^{n})+I(M_{1},M_{2};X_{1}^{n},X_{2}^{n}|A^{n},Y^{n})\nonumber \\
 & \overset{}{=}I(A^{n};X_{1}^{n},X_{2}^{n})+H(X_{1}^{n},X_{2}^{n}|A^{n},Y^{n})-H(X_{1}^{n},X_{2}^{n}|A^{n},Y^{n},M_{1},M_{2})\nonumber \\
 & \overset{}{=}H(X_{1}^{n},X_{2}^{n})-H(X_{1}^{n},X_{2}^{n}|A^{n})+H(X_{1}^{n},X_{2}^{n},Y^{n}|A^{n})-H(Y^{n}|A^{n})\nonumber \\
 & -H(X_{2}^{n}|A^{n},Y^{n},M_{1},M_{2})-H(X_{1}^{n}|A^{n},Y^{n},X_{2}^{n},M_{1},M_{2})\nonumber \\
 & \overset{(b)}{\geq}H(X_{1}^{n},X_{2}^{n})+H(Y^{n}|A^{n},X_{1}^{n},X_{2}^{n})-H(Y^{n}|A^{n})\nonumber \\
 & -H(X_{1}^{n}|A^{n},Y^{n},X_{2}^{n},M_{1},M_{2})-n\delta(\epsilon),\label{eq3-1}
\end{align}
where (\textit{a}) follows because $A^{n}$ is a function of $M_{1}$;
and (\emph{b}) follows as in ($a$) of (\ref{R2_conv_opt2}). For
the first three terms in (\ref{eq3-1}) we have
\begin{align}
 & H(X_{1}^{n},X_{2}^{n})+H(Y^{n}|A^{n},X_{1}^{n},X_{2}^{n})-H(Y^{n}|A^{n})\nonumber \\
 & \overset{(a)}{=}\overset{n}{\underset{i=1}{\sum}}H(X_{1i},X_{2i})+H(Y_{i}|Y^{i-1},A^{n},X_{1}^{n},X_{2}^{n})-H(Y_{i}|Y^{i-1},A^{n})\nonumber \\
 & \overset{(b)}{\geq}\overset{n}{\underset{i=1}{\sum}}H(X_{1i},X_{2i})+H(Y_{i}|A_{i},X_{1i},X_{2i})-H(Y_{i}|A_{i})\nonumber \\
 & \overset{}{=}\overset{n}{\underset{i=1}{\sum}}H(X_{1i},X_{2i})-I(Y_{i};X_{1i},X_{2i}|A_{i})\nonumber \\
 & \overset{}{=}\overset{n}{\underset{i=1}{\sum}}H(X_{1i},X_{2i})-H(X_{1i},X_{2i}|A_{i})+H(X_{1i},X_{2i}|A_{i},Y_{i})\nonumber \\
 & \overset{}{=}\overset{n}{\underset{i=1}{\sum}}I(X_{1i},X_{2i};A_{i})+H(X_{2i}|A_{i},Y_{i})+H(X_{1i}|A_{i},Y_{i},X_{2i}),\label{eq4-1}
\end{align}
\\
where (\textit{a}) follows from the chain rule for entropy and by
the chain rule for entropy and the fact that $(X_{1}^{n},X_{2}^{n})$
are i.i.d.; and (\textit{b}) follows since $Y_{i}$---$(A_{i},X_{\{1,2\}i})$---$(Y^{i-1},A^{n\backslash i},X_{1}^{n\backslash i},X_{2}^{n\backslash i})$
forms a Markov chain, by the definition of problem, and since conditioning
reduces entropy. Combining (\ref{eq3-1}) and (\ref{eq4-1}), and
using the definition of $U_{1i}$, we obtain
\begin{align}
 & n(R_{1}+R_{2})\overset{(a)}{\geq}\overset{n}{\underset{i=1}{\sum}}I(X_{1i},X_{2i};A_{i})+H(X_{2i}|A_{i},Y_{i})+H(X_{1i}|A_{i},Y_{i},X_{2i})\nonumber \\
 & -H(X_{1i}|X_{1}^{i-1},A^{n},Y^{n},X_{2}^{n},M_{1},M_{2})-n\delta(\epsilon)\nonumber \\
 & \overset{(b)}{\geq}\overset{n}{\underset{i=1}{\sum}}I(X_{1i};A_{i})+H(X_{2i}|A_{i},Y_{i})+H(X_{1i}|A_{i},Y_{i},X_{2i})\nonumber \\
 & -H(X_{1i}|A^{n},Y^{n},X_{2}^{n},M_{1})-n\delta(\epsilon)\nonumber \\
 & \overset{(c)}{\geq}\overset{n}{\underset{i=1}{\sum}}(X_{1i};A_{i})+H(X_{2i}|A_{i},Y_{i})+I(X_{1i};U_{1i}|A_{i},Y_{i},X_{2i})-n\delta(\epsilon),\label{R1+R2_conv_opt2}
\end{align}
where (\textit{a}) follows by the chain rule for entropy; (\textit{b})
follows because mutual information is non-negative and due to the
fact that conditioning decreases entropy; and (\textit{c}) follows
by the definition of mutual information and definition of $U_{1i}$
and the fact that conditioning decreases entropy.

\begin{figure}[h!]
\centering\includegraphics[bb=88bp 427bp 417bp 707bp,clip,scale=0.75]{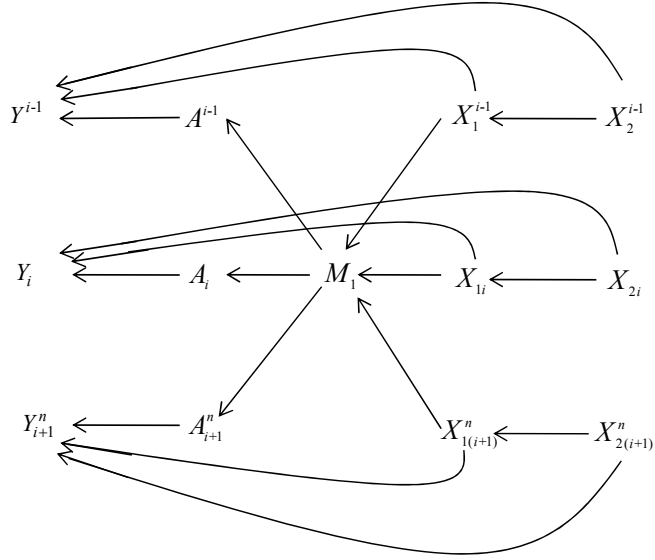}
\caption{Bayesian network representing the joint pmf of variables ($M_{1},X_{1}^{n},X_{2}^{n},A^{n},Y^{n}$)
for the model in Fig. \ref{fig:fig1}.}

\label{fig:graph1} 
\end{figure}

Moreover, $(X_{2i},Y_{i})-(X_{1i},A_{i})-U_{1i}$ forms a Markov chain.
This can be seen by using the principle of $d$-separation \cite[Sec. A.9]{Kramer}
from Fig. \ref{fig:graph1}, which represents the joint distribution
of all the variables at hand.

Let $Q$ be a uniform random variable over the interval $[1,n]$ and
independent of $(X_{1}^{n},X_{2}^{n},A^{n},$
 $U_{1}^{n},Y^{n},\hat{X}_{1}^{n})$ and define $U_{1}\overset{\Delta}{=}(Q,U_{1Q})$,
$X_{1}\overset{\Delta}{=}X_{1Q}$, $X_{2}\overset{\Delta}{=}X_{2Q}$,
$Y\overset{\Delta}{=}Y_{Q}$, $A\overset{\Delta}{=}A_{Q}$, and $\hat{X}_{1}\overset{\Delta}{=}\hat{X}_{1Q}.$
Note that $\hat{X}_{1}$ is a function of $U_{1}$ and $Y$. Moreover,
from (\ref{action_cost_const}) and (\ref{eq:dist_const}), we have
\begin{align}
\Gamma+\epsilon & \geq\frac{1}{n}\underset{i=1}{\overset{n}{\sum}}\mathrm{E}\left[\Lambda(A_{i})\right]=\mathrm{E}[\Lambda(A)]\nonumber \\
\text{and }D_{1}+\epsilon & \geq\frac{1}{n}\underset{i=1}{\overset{n}{\sum}}\mathrm{E}\left[d_{1}(X_{1i},X_{2i},Y_{i}.\hat{X}_{1i})\right]=\mathrm{E}[d_{1}(X_{1},X_{2},Y,\hat{X}_{1})].
\end{align}

Finally, since (\ref{R1_conv_opt2}), (\ref{R2_conv_opt2}) and (\ref{R1+R2_conv_opt2})
are convex with respect to $p(a,u_{1}|x_{1},q)$ for fixed $p(q)$,
$p(x_{1},x_{2})$, and $p(y|a,x_{1},x_{2})$, we have that inequalities
(\ref{DSC_opt2}) hold, which completes the proof of (\ref{R1_opt2})-(\ref{eq:dist_const_opt2}).
The cardinality bounds are proved by using the Fenchel--Eggleston--Caratheodory
theorem in the standard way.

\section*{Appendix D}

\section*{Greedy Actions Are Optimal With Sum Side Information}

Here we prove equality

\begin{equation}
R_{sum,\text{ }greedy}^{\oplus}(\Gamma)=R_{sum}^{\oplus}(\Gamma).\label{equality greedy}
\end{equation}
which shows that no gain is accrued by choosing the actions based
only on message $M_{1}$ with the sum side information. Fix the pmf
$p(a|x_{1})$ that achieves the minimum in the sum-rate obtained from
(\ref{R1+R2_opt2}), namely
\begin{eqnarray*}
R_{sum}^{\oplus}(\Gamma) & = & \textrm{min }I(X;A)+H(X_{1},X_{2}|A,Y),
\end{eqnarray*}
where the mutual information is calculated with respect to the distribution
\begin{equation}
p(x_{1},x_{2},y,a)=p(x_{1},x_{2})p(a|x_{1})p(y|a,x_{1},x_{2}),\label{Rsum_joint}
\end{equation}
and the minimum is taken over all distributions $p(a|x_{1})$ such
that $\mathrm{E}\left[\Lambda(A)\right]=\mathrm{E}\left[A\right]\leq\Gamma.$
Note that for such a pmf $p(a|x_{1})$ we have $\mathrm{E}[A]=p(a)=\Gamma,$
as it can be easily seen. We then have the following series of equalities:
\begin{align*}
 & R_{sum,\text{ }greedy}^{\oplus}(\Gamma)-R_{sum}^{\oplus}(\Gamma)\\
 & \overset{(a)}{=}\Gamma H(X_{1},X_{2}|X_{1}\oplus X_{2})+(1-\Gamma)H(X_{1},X_{2})\\
 & -H(X_{1},X_{2}|A,X_{1}\oplus X_{2})-I(X_{1};A)\\
 & \overset{(b)}{=}\Gamma H(X_{1}|X_{1}\oplus X_{2})+(1-\Gamma)(1+H(p))-\Gamma H(X_{1},X_{2}|A=1,X_{1}\oplus X_{2})\\
 & -(1-\Gamma)H(X_{1},X_{2}|A=0)-I(X_{1};A)\\
 & \overset{(c)}{=}\Gamma H(X_{1})+(1-\Gamma)(1+H(p))-\Gamma H(X_{1}|A=1)-(1-\Gamma)H(X_{1}|A=0)\\
 & -(1-\Gamma)H(X_{2}|X_{1},A=0)-I(X_{1};A)\\
 & \overset{(d)}{=}\Gamma+(1-\Gamma)(1+H(p))-H(X_{1}|A)-(1-\Gamma)H(X_{2}|X_{1})-I(X_{1};A)\\
 & =\Gamma+(1-\Gamma)(1+H(p))-H(X_{1}|A)-(1-\Gamma)H(p)-1+H(X_{1}|A)=0,
\end{align*}
where (\textit{a}) follows by the definition (\ref{Rsumgreedy});
(\textit{b}) follows using the chain rule for entropy and from the
definition of conditional entropy; (\textit{c}) follows by the crypto-lemma
\cite[Lemma 2]{Forney}; (\textit{d}) follows from the fact that $X_{2}-X_{1}-A$
forms a Markov chain.

\section*{Appendix E}

\section*{Proof of the Converse for Proposition \ref{prop:3}}

In this section, we provide the proof of converse for Proposition
\ref{prop:3}. For any $(n,R_{12},R_{23},D_{1}+\epsilon,D_{2}+\epsilon,\Gamma+\epsilon)$
code, we have the following inequalities:
\begin{align}
nR_{12} & \geq H(M_{12})\overset{}{\geq}H(M_{12}|X_{2}^{n})\overset{(a)}{=}H(M_{12},M_{23}|X_{2}^{n})\nonumber \\
 & \overset{(b)}{=}I(X_{1}^{n};M_{12},M_{23}|X_{2}^{n})\nonumber \\
 & \overset{}{=}\underset{i=1}{\overset{n}{\sum}}H(X_{1i}|X_{1}^{i-1},X_{2}^{n})-H(X_{1i}|X_{1}^{i-1},X_{2}^{n},M_{12},M_{23})\nonumber \\
 & \overset{(c)}{=}\underset{i=1}{\overset{n}{\sum}}H(X_{1i}|X_{2i})-H(X_{1i}|X_{1}^{i-1},X_{2}^{n},A^{n},M_{12},M_{23})\nonumber \\
 & \overset{(d)}{=}\underset{i=1}{\overset{n}{\sum}}H(X_{1i}|X_{2i})-H(X_{1i}|X_{1}^{i-1},X_{2}^{n},Y^{i-1},M_{12},M_{23},A^{n},\hat{X}_{1}^{n})\nonumber \\
 & \overset{(e)}{\geq}\underset{i=1}{\overset{n}{\sum}}H(X_{1i}|X_{2i})-H(X_{1i}|X_{2i},A_{i},U_{i},\hat{X}_{1i})\nonumber \\
 & \overset{}{=}\underset{i=1}{\overset{n}{\sum}}I(X_{1i};A_{i},U_{i},\hat{X}_{1i}|X_{2i}),\label{R1}
\end{align}
where (\textit{a}) follows because $M_{23}$ is a function of ($M_{12}$,$X_{2}^{n})$;
(\emph{b}) follows by definition of mutual information and since $M_{12}$
and $M_{23}$ are functions of $X_{1}^{n}$ and $X_{2}^{n}$; (\textit{c})
follows because $X_{1}^{n}$ and $X_{2}^{n}$ are i.i.d and since
$A^{n}$ is a function of $M_{23};$ (\emph{d}) follows because $Y^{i-1}-(X_{1}^{i-1},X_{2}^{n},A^{n},M_{12},M_{23})-X_{1i}$
forms a Markov chain and since $\hat{X}_{1}^{n}$ is a function of
$M_{12}$ and $X_{2}^{n}$; and (\emph{e}) follows by defining $U_{i}=(X_{1}^{i-1},X_{2}^{i-1},Y^{i-1},A^{n\backslash i},M_{23})$
and since conditioning decreases entropy.

We also have the inequalities
\begin{align}
nR_{23} & \geq H(M_{23})\overset{(a)}{=}I(X_{1}^{n},X_{2}^{n};M_{23})\nonumber \\
 & \overset{(b)}{=}\underset{i=1}{\overset{n}{\sum}}H(X_{1i},X_{2i})-H(X_{1i},X_{2i}|X_{1}^{i-1},X_{2}^{i-1},M_{23})\nonumber \\
 & \overset{(c)}{=}\underset{i=1}{\overset{n}{\sum}}H(X_{1i},X_{2i})-H(X_{1i},X_{2i}|X_{1}^{i-1},X_{2}^{i-1},A^{n},M_{23})\nonumber \\
 & \overset{(d)}{=}\underset{i=1}{\overset{n}{\sum}}H(X_{1i},X_{2i})-H(X_{1i},X_{2i}|X_{1}^{i-1},X_{2}^{i-1},Y^{i-1},A^{n},M_{23})\nonumber \\
 & \overset{(e)}{=}\underset{i=1}{\overset{n}{\sum}}H(X_{1i},X_{2i})-H(X_{1i},X_{2i}|A_{i},U_{i})\nonumber \\
 & \overset{}{=}\underset{i=1}{\overset{n}{\sum}}I(X_{1i},X_{2i};A_{i},U_{i}),\label{R2}
\end{align}
where (\textit{a}) follows because $M_{23}$ is a function of $X_{1}^{n}$
and $X_{2}^{n}$; (\textit{b}) follows by the definition of mutual
information and the chain rule for entropy and since $X_{1}^{n}$
and $X_{2}^{n}$ are i.i.d; (\textit{c}) follows because $A^{n}$
is a function of $M_{23}$; (\textit{d}) follows because $Y^{i-1}-(X_{1}^{i-1},X_{2}^{i-1},A^{n},M_{23})-(X_{1i},X_{2i})$
forms a Markov chain; and (\textit{e}) follows by the definition of
$U_{i}$.

Let $Q$ be a uniform random variable over $[1,n]$ and independent
of $(X_{1}^{n},X_{2}^{n},Y^{n},A^{n},U^{n},\hat{X}_{1}^{n})$ and
define $U\overset{\Delta}{=}(Q,U_{Q})$, $X_{1}\overset{\Delta}{=}X_{1Q}$,
$X_{2}\overset{\Delta}{=}X_{2Q}$, $Y\overset{\Delta}{=}Y_{Q}$, $A\overset{\Delta}{=}A_{Q}$,
$\hat{X}_{1}\overset{\Delta}{=}\hat{X}_{1Q},$ and $\hat{X}_{2}\overset{\Delta}{=}\hat{X}_{2Q}$.
Note that $\hat{X}_{2}$ is a function of $U$ and $Y$. Moreover,
from (\ref{action cost}) and (\ref{dist const}), we have
\begin{align}
\Gamma+\epsilon & \geq\frac{1}{n}\underset{i=1}{\overset{n}{\sum}}\mathrm{E}\left[\Lambda(A_{i})\right]=\mathrm{E}[\Lambda(A)]\label{action_dist}\\
\text{and }D_{j}+\epsilon & \geq\frac{1}{n}\underset{i=1}{\overset{n}{\sum}}\mathrm{E}\left[d_{j}(X_{1i},X_{2i},Y_{i},\hat{X}_{ji})\right]=\mathrm{E}[d_{j}(X_{1},X_{2},Y,\hat{X}_{j})]\text{ for }j=1,2.
\end{align}

Finally, since (\ref{R1}) and (\ref{R2}) are convex with respect
to $p(a,u,\hat{x}_{1}|x_{1},x_{2})$ for fixed $p(x_{1},x_{2})$ and
$p(y|a,x_{1},x_{2})$, we have from (\ref{R1}) and (\ref{R2}) that
inequalities (\ref{reg_cascade}) hold. The cardinality bounds are
proved by using the Fenchel\textendash{}Eggleston\textendash{}Caratheodory
theorem in the standard way.

\end{document}